\documentclass[aps,preprintnumbers,11pt,amsmath,amssymb,nofootinbib]{revtex4}

\usepackage{braket}
\usepackage{float}
\usepackage{graphics,epsfig}
\usepackage{bm}
\usepackage{slashed}
\usepackage{graphicx}
\usepackage{amsmath}
\usepackage{amsfonts}
\usepackage{amssymb}
\usepackage{color}
\usepackage{dcolumn}
\usepackage[
            pdfstartview=FitH,
            bookmarksnumbered=true,
            bookmarksopen=true,
            colorlinks,
            linkcolor=blue,
            anchorcolor=green,
            citecolor=red
            ]{hyperref}
\usepackage{enumerate}
\providecommand{\U}[1]{\protect\rule{.1in}{.1in}}

 \renewcommand\arraystretch{2}
 \newcommand{\bq}{\begin{equation}}
 \newcommand{\eq}{\end{equation}}
 \newcommand{\bqn}{\begin{eqnarray}}
 \newcommand{\eqn}{\end{eqnarray}}
 \newcommand{\nb}{\nonumber}
 \newcommand{\lb}{\label}
 
\def\A0{A^{(0)}}

\begin{document}
\baselineskip=0.6 cm \title{Extended geometry of Gambini-Olmedo-Pullin polymer black hole and its quasinormal spectrum}

\author{Yu-Chen Liu$^{1,2}$}
\thanks{liuyuchenloop@gmail.com}
\author{Jia-Xi Feng $^{1,2}$}
\thanks{fengjiaxigw@gmail.com}
\author{Fu-Wen Shu$^{1,2}$}
\thanks{shufuwen@ncu.edu.cn; the corresponding author}
\author{Anzhong Wang$^{3}$}
\thanks{Anzhong$\_$Wang@baylor.edu}

\affiliation{
$^{1}$Department of Physics, Nanchang University, Nanchang, 330031, China\\
$^{2}$Center for Relativistic Astrophysics and High Energy Physics, Nanchang University, Nanchang 330031, China\\
$^{3}$GCAP-CASPER, Physics Department, Baylor University, Waco, Texas 76798-7316, USA}

\vspace*{0.2cm}

\vspace*{0.2cm}
\begin{abstract}
\baselineskip=0.6 cm
\begin{center}
{\bf Abstract}
\end{center}
In this paper we systematically study a model of spherically symmetric polymer black holes recently proposed by Gambini, Olmedo, and Pullin (GOP). Within the framework of loop quantum gravity, the quantum parameters in the GOP model depend on the minimal area gap and the size of the discretization of the physical states. In this model, a spacelike transition surface takes the place of the classical singularity. By means of coordinate transformations, we first extend the metric to the white hole region, and find that the geometric structure of the quantum black hole is similar to the wormhole structure, and the radius of the most quantum region is equal to the wormhole radius. In addition, we show that the energy conditions are violated not only at the throat but also at the horizons and the spatial infinities. In order to show how the quantum effects affect the spacetimes, we calculate the Ricci and Kretschmann scalars at different places. It turns out that, as expected, the most quantum region is at the throat. Finally, we consider the quasinormal modes (QNMs) of massless scalar field perturbations, electromagnetic field perturbations, and axial gravitational perturbations. QNMs in the Eikonal limits are also considered. As anticipated, the spectrum of QNMs deviates from that of the classical case due to quantum effects. Interestingly, our results show that the quasinormal frequencies of the perturbations share the same qualitative tendency while setting quantum parameters with  various values in this effective model, even if the potential deviations are different with different spins.

\end{abstract}

\maketitle
\newpage
\vspace*{0.2cm}

\section{Introduction}
There is much evidence that show that the predictions of general relativity (GR) are not reliable and the effects of quantum gravity would take over once the curvature of spacetime enters the Planck regime. Among them spacetime singularities are of particular significance and hence have attracted considerable attention over the past few decades.  It is widely believed that classical singularities should be resolved properly under the framework of quantum theory of gravity. Since a complete and consistent quantum theory of gravity is still missing, in the past decades, efforts towards the understanding of the spacetime singularities have been mostly made to effective alternatives. One of the most successful examples is the application of an effective approach developed in loop quantum gravity (LQG) to the big bang singularity in cosmology \cite{Singh:2009mz,Ashtekar:2011ni}. More recently, attempts to extend the approaches developed in
loop quantum cosmology to black hole singularities have been investigated intensively \cite{Ashtekar:2005qt,Modesto:2005zm,Boehmer:2007ket,Campiglia:2007pb,Brannlund:2008iw,Modesto:2008im,Chiou:2008nm,Chiou:2008eg,GP08,Gambini:2013exa,Joe:2014tca,Corichi:2015xia,Olmedo:2017lvt,Cortez:2017alh,CR17,AP17,BMM18,RMD18,BCDHR18,AOS18a,AOS18b,Alesci:2019pbs,Assanioussi:2019twp,Bodendorfer:2019nvy,MDR19,AAN20,Ashtekar:2020ckv,Zhang:2020qxw,Gambini:2020nsf,Kelly:2020uwj,Liu:2020ola,Agullo:2020hxe,Giesel:2021dug,Garcia-Quismondo:2021xdc}. In all these situations a phase space regularization called polymerization plays a key role \cite{Thiemann08}. Motivated by a mini-superspace polymerlike quantization \cite{Ashtekar:2002sn,Ashtekar:2003hd,Corichi:2007tf}  inspired by LQG, one replaces the canonical momenta of the theory with their holonomies, an exponentiated version of the canonical variables including two quantum parameters $\delta_b$ and $\delta_c$ for spherical spacetimes \cite{Ashtekar20}. More precisely,  the effective quantum theory can be achieved by replacing the canonical variables ($b, c$) in the phase space with their regularized ones,
\bqn
\lb{eq1}
b &\rightarrow&\frac{\sin(\delta_b b)}{\delta_b},\quad c \rightarrow\frac{\sin(\delta_c c)}{\delta_c},
\eqn
where $\delta_b$ and $\delta_c$ are called  ``polymerization scales ,'' which control the onset of quantum effects. As $\delta_b$ and $\delta_c$ go to $0$, the effective Hamiltonian reduces to a classical one, indicating the classical limit is recovered in this limit. On the other hand,  as quantum effects become relevant, where $\delta_b$ and $\delta_c$  are comparable with the Planck scale, the classical divergence can be effectively avoided through the replacement \eqref{eq1}.

However, a full picture on how to choose $\delta_b$ and $ \delta_c$ is still missing, since a complete theory of quantum gravity is still lacking.  Nonetheless, over the past few years many different choices have been proposed. Initially, $\mu_0$ scheme was proposed  \cite{Ashtekar:2005qt, Bodendorfer:2019nvy, Assanioussi:2019twp, Campiglia:2007pb, Modesto:2005zm, Modesto:2008im}, in which these two quantum parameters $\delta_b$ and $ \delta_c$ are simply set to constants. However, there is a significant drawback in this scheme: quantum effects domination occurs at an arbitrarily low curvatures scale, which is obviously in contradiction with the facts. Soon after, it was found that this limitation can be resolved if the polymerization parameters depend on the canonical variables, that is what we called $\overline{\mu}$ scheme (or improved dynamics) \cite{Chiou:2008nm,Boehmer:2007ket,Brannlund:2008iw,Cortez:2017alh,Chiou:2008eg,Joe:2014tca,Alesci:2019pbs}. There is also an approach in between,  the generalized $\mu_0$ scheme\cite{Ashtekar:2020ckv, Corichi:2015xia, Olmedo:2017lvt,AOS18a,AOS18b}, where the quantum parameters $\delta_b$ and $\delta_c$ are considered as the Dirac observables. That is to say, they are constants along the effective trajectories of the system. For other cases, they are generally phase space functions. Later, a variant of $\overline{\mu}$ scheme was proposed in \cite{BMM19,BMM20},  where the authors introduced a new classical phase space description based on canonical variables inspired by physical considerations about the onset of quantum effects. It turns out many desirable features of the resulting
quantum corrected spacetime can be obtained \cite{BMM19,BMM20,Bouhmadi-Lopez:2020oia,Gan:2020dkb}.

Very recently, an improved quantization scheme for spherically symmetric loop quantum gravity was proposed by Gambini, Olmedo, and Pullin (GOP) in \cite{Gambini:2020nsf,Gambini:2020qhx}. In particular, they first consider a kinematical Hilbert space in the loop representation adapted to spherically symmetric spacetimes with geometric triad variables $(E^{\varphi}, E^{x})$ and their conjugates $(K_{\varphi}, K_x)$, and a representation for the spacetime mass and its conjugate momentum as well. Then, they represent the scalar constraint as an operator in the kinematical Hilbert space. By applying group averaging techniques for both the quantum scalar constraint and the group of finite spatial diffeomorphisms, one can obtain the physical states $|M,\vec k\rangle$, where $M$ labels the Arnowitt-Deser-Misner (ADM) mass of the spin network and $k_i\in Z$ are valences of edges of the network. Finally, using some parametrized observables that act as local operators on each vertex of the spin network, one can define the physical observables denoting space-time metric components.

In this paper, we will focus on this effective quantum black hole (the GOP black hole for short) \cite{Gambini:2020nsf}.  In the previous literature, although some geometrical properties have been explored, the full understanding of its structure is still absent. For instance, the knowledge of what is the other side of the transition surface is lacking. To achieve this, following \cite{Gambini:2020nsf,Gambini:2020qhx} we first diagonalize the metric and extend it to the region $x<0$ which can be referred to as a white hole region.
We find that the black hole and the white hole near the transition surface cannot connect smoothly.  We then investigate the main properties of the quantum black hole. In particular, following \cite{Bouhmadi-Lopez:2020oia,Gan:2020dkb}, we examine the energy conditions by treating the quantum corrections on the spacetime as a kind of effective matter field. We find that the energy conditions are violated as expected in the full parameter space in the whole spacetimes. In order to show how the effects of quantum gravity affect the geometry, we also explore the departure of the effective metric from classical GR at different regions of the spacetimes by calculating the Ricci and Kretschmann scalars at each region. As the last part, we study the perturbations of the GOP quantum black hole and calculate their quasinormal mode (QNM) frequencies for three different cases: the massless scalar field perturbations, the electromagnetic field perturbations, and the axial gravitational perturbations. We compare our results with those of the Schwarzschild black hole and demonstrate how the QNM frequencies change with the quantum parameters. We find that the quasinormal frequencies of the perturbation share the same qualitative tendency while setting quantum parameters with  various values in this effective model, even if the potential deviations are different with different spins.

This paper is outlined as follows. In Sec. II, we study the extension of the GOP black hole. In Sec. III, we discuss physical properties of the quantum black hole, including energy conditions and quantum deviations of the curvature scalars. In Sec. IV, we study the massless scalar field perturbations, the electromagnetic field perturbations, and the axial gravitational perturbations of the quantum black hole. QNMs in the Eikonal limit are also discussed in this section. A brief concluding remark is drawn in Sec. V.

\section{Extended geometry of GOP black hole}
The most general metric for a spherically symmetric spacetime is given by
\bq\label{metric1}
ds^2=-(N^2-N_xN^x)dt^2+2N_xdtdx+\frac{(E^\varphi)^2}{|E^x|}dx^2+|E^x|d\omega^2,
\eq
where $N$ is the lapse function, $N_x$ is the shift vector, $E^{\varphi}$ and $E^x$ are triad variables, and $d\omega^2$ is the line element of a unit 2-sphere.
If we make a transformation in the following way \cite{Gambini:2020qhx}
\bqn
x&\rightarrow&x,\nonumber\\
t&\rightarrow&t+\int dx\frac{N_x}{N^2-N_xN^x},
\eqn
we then find that the metric takes a simpler form,
\begin{equation}\label{metric2}
ds^2=-(N^2-N_xN^x)dt^2+\left(\frac{(N_x)^2}{N^2-N_xN^x}+\frac{(E^\varphi)^2}{|E^x|}\right)dx^2+|E^x|d\omega^2.
\end{equation}
For the GOP polymer black hole \cite{Gambini:2020nsf},  we have
\bqn
\label{d}
N^2-N_xN^x&=&1-\frac{r_S}{x+x_0}+\frac{\Delta}{4\pi}\frac{r_S^4 }{(x+x_0)^6\left(1+\frac{r_S}{x+x_0}\right)^2},\nonumber\\
2N_x&=&2\frac{r_S}{(x+x_0)}\left(1+\frac{\delta x}{2(x+x_0)}\right)\left(\sqrt{1-\frac{\Delta}{4\pi}\frac{r_S^2 }{(x+x_0)^4\left(1+\frac{r_S}{x+x_0}\right)}}\right),\nonumber\\
\frac{(E^\varphi)^2}{|E^x|}&=&\left(1+\frac{r_S}{x+x_0}\right)\left(1+\frac{\delta x}{2(x+x_0)}\right)^2,\nonumber\\
|E^x|&=&(x+x_0)^2,
\eqn
where $x\geq0$ and $r_S=2GM_0$ is the classical Schwarzschild radius, $x_0=\left(\frac{2GM_0\Delta}{4\pi}\right)^{\frac{1}{3}}$ represents a scale below which quantum effects cannot be ignored,\footnote{The above metric agrees well with the classical metric when $x\gtrsim x_0$, while the quantum effect plays a dominant role when $x<x_0$ and spacetimes enter into the high curvature region.} $\Delta$ is the area gap parameter, and $\delta x$ is the step of the lattice of the coordinate $x$ and it is generally chosen to be $\delta x=\ell_{\rm Pl}$. \footnote{ In what follows, we will adopt the natural units $\hbar=c=G=1$, which means that $\ell_{\rm Pl}^2=1$.}

For simplicity, we introduce a new variable
\begin{equation}
\label{b}
X:=\frac{r_S}{x+x_0}, \ \  x\geq0,
\end{equation}
and define two dimensionless parameters of the theory
\bq
\label{c}
\alpha:=\frac{\Delta}{4\pi r_S^2},\ \
\beta:=\frac{\delta x}{2r_S}.
\eq
The metric then can be cast as
\begin{equation}
\label{metric3}
ds_+^2=-a(X)dt^2+b(X)dX^2+r^2(X)d\omega^2,
\end{equation}
where we have defined the following functions:
\bqn
a(X)&=&1-X+\frac{\alpha X^6}{(1+X)^2},\label{ax}\\
b(X) &=& \frac{\Delta(1+X)(1+\beta X)^2}{4\pi\alpha X^4}\left(\frac{X^2(1+X-\alpha X^4)}{(1-X)(1+X)^2+\alpha X^6}+1\right),\\
r(X)&=& \sqrt{\frac{\Delta}{4\pi\alpha}} \frac{1}{X}\label{rx}.
\eqn

Note that the above metric is only valid for $x\geq0$. However, it can be extended to $x<0$ by letting $x\rightarrow-x$, $x_0\rightarrow-x_0$, $r_S\rightarrow-r_S$, and $\delta x\rightarrow-\delta x$. Actually, all the metric components \eqref{d}, $\alpha$ and $\beta$ as well, are invariant under this transformation. Hence, the theory admits the same solution \eqref{metric3}--\eqref{rx} but now with
\begin{equation}
\label{X-} X=\frac{r_S}{x_0-x},\ \ \ \text{for} \ x<0.
\end{equation}
We should emphasize that although the solution has the same form as \eqref{metric3}--\eqref{rx}, it belongs to a different branch (the $x<0$ branch). To avoid any confusion, let us denote this branch by $ds_-^2$ in the following discussion.

It is also helpful to know that this extension is equal to the GOP extension with $x\rightarrow|x|$ which is very similar to the extension made in \cite{Gambini:2020qhx}. In other words, after the extension, the theory admits the solution \eqref{metric3}--\eqref{rx} with
\begin{equation}
\label{X} X=\frac{r_S}{x_0+|x|},\ \ \ \text{for} \ x\in(-\infty,\infty).
\end{equation}

\begin{figure}[h!]
{\centering
  \includegraphics[width = 0.5\textwidth]{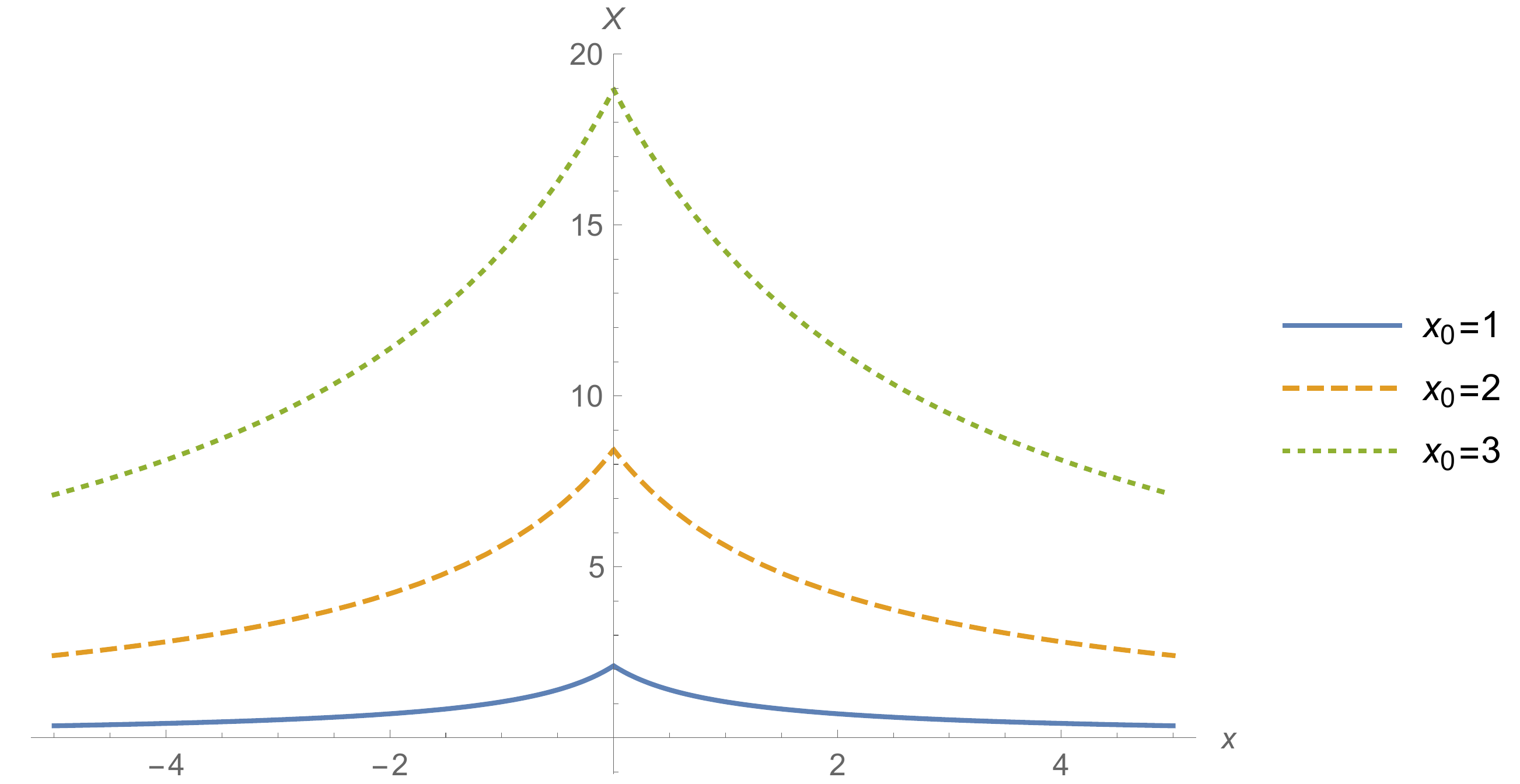}
}
\caption{The parameter $X$ as a function of $x$: When plotting these curves, the solid, dashed, and dotted curves correspond to $x_0=1$, $x_0=2$, and $x_0=3$, respectively.}
\label{figXx}
\end{figure}

\begin{figure}[h!]
{\centering
  \includegraphics[width = 0.5\textwidth]{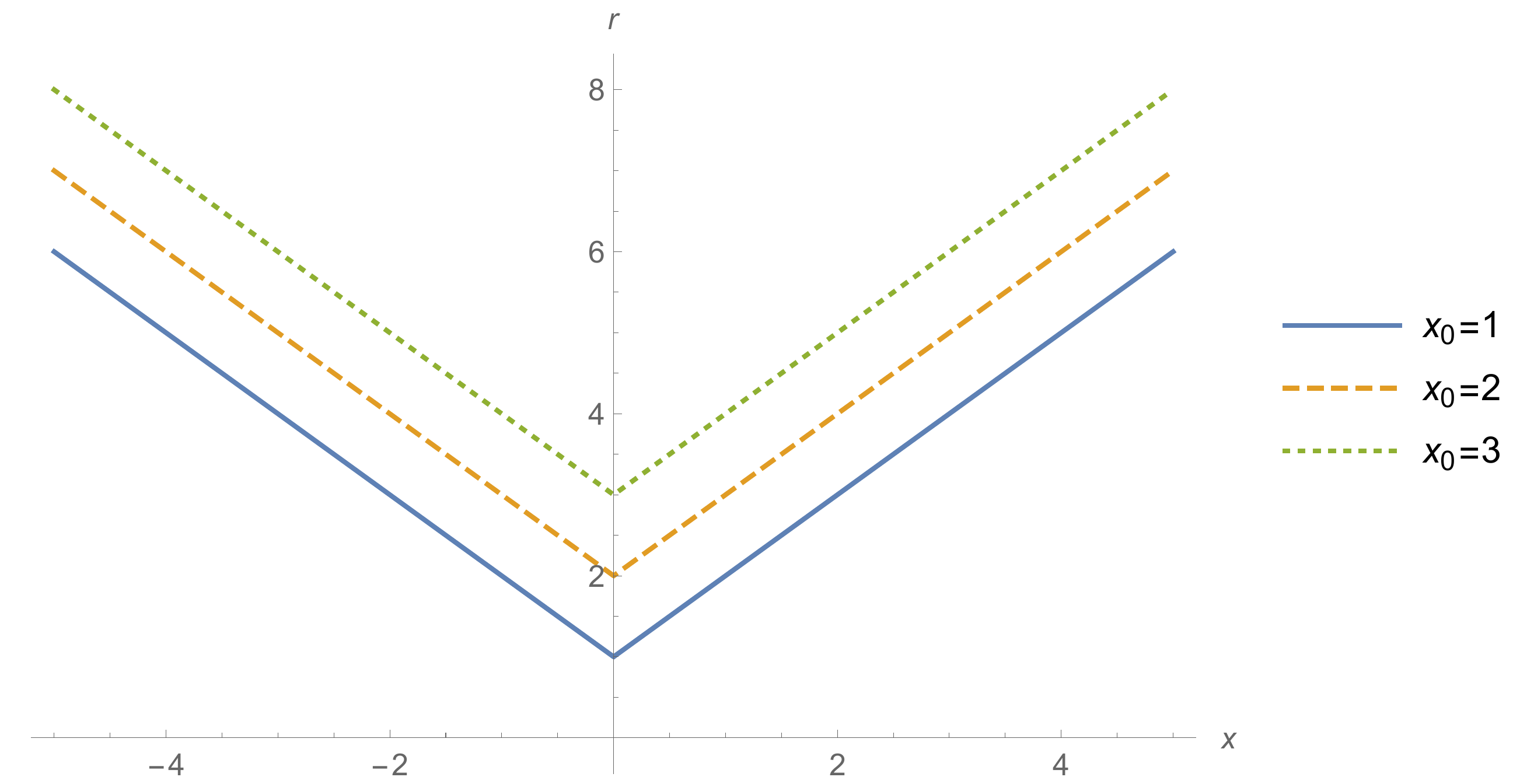}
}
\caption{The geometric radius $r(X)$ as a function of $x$:  When plotting these curves,  the solid, dashed, and dotted curves correspond to $x_0=1$, $x_0=2$, and $x_0=3$ respectively.}
\label{figrx}
\end{figure}

\begin{figure}[h!]
{\centering
  \includegraphics[width = 0.35\textwidth]{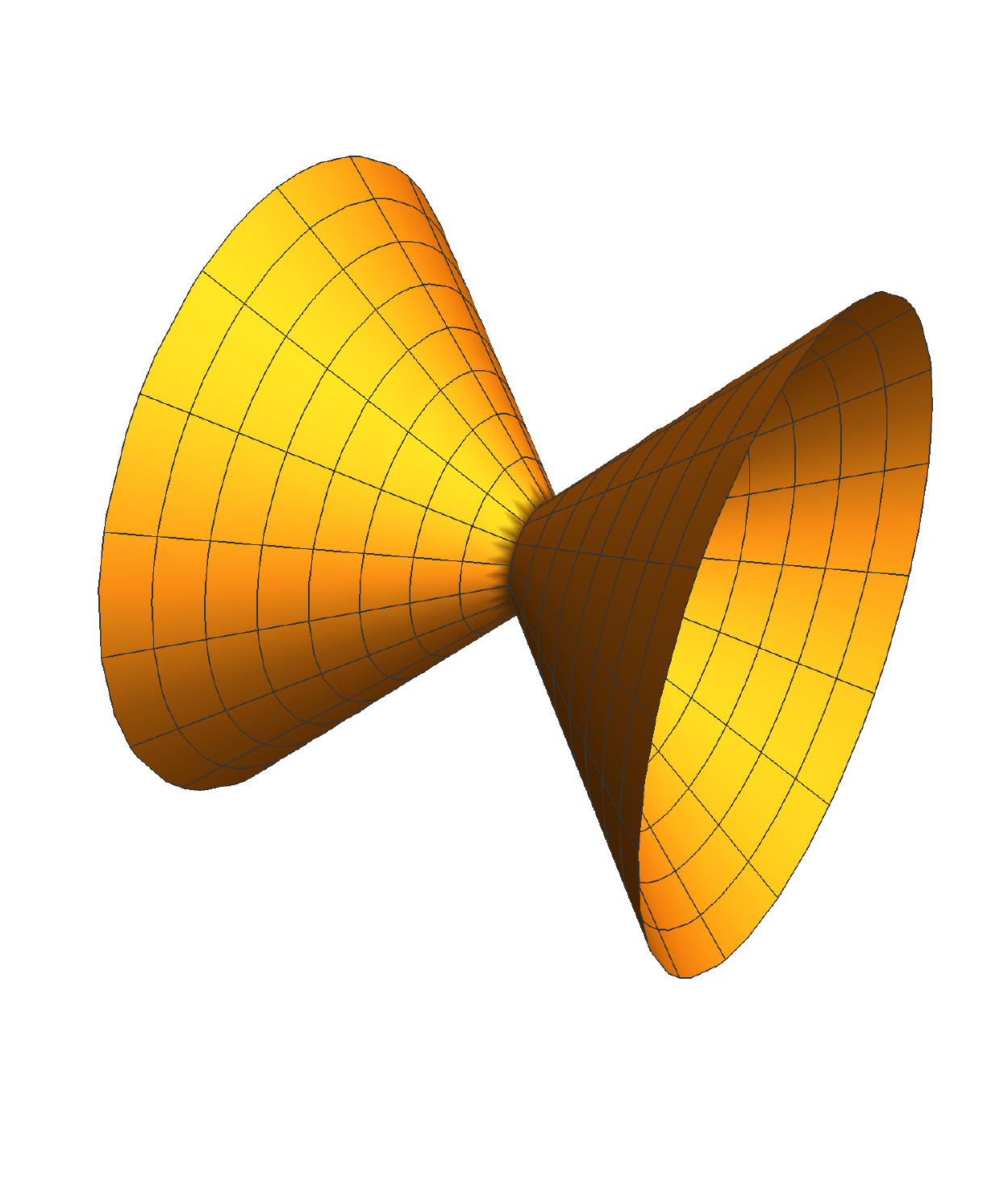}
}
\caption{The sketch of the wormhole.}
\label{throat}
\end{figure}

Notice that $X$ is always positive regardless if $x>0$ or $x<0$,\footnote{Precisely, $X$ can only take the value bigger than $\frac{r_S}{x_0}$ as shown in \eqref{X}. It is in this sense we say that the metric \eqref{metric3}--\eqref{rx} defined in terms of $X$ is incomplete. We treat the $X$-frame as an auxiliary frame, instead of the physical one.} as shown in Fig. \ref{figXx}. As a consequence, the geometric radius $r(x)$ varies linearly with $x$, as shown in Fig. \ref{figrx}. As $x\rightarrow 0$, the geometric radius reaches its minimum $r_{T}=\sqrt{\frac{\Delta}{4\pi\alpha}} \frac{x_0}{r_S}=x_0$. Hence, the spacetime forms a wormhole, with the throat located at $x=0$, as sketched in Fig. \ref{throat}.

After performing the above extension, due to time reversal symmetry,  the $x>0$ branch can be identified as a black hole, while the $x<0$ branch is viewed as a white hole. In addition, one can show the asymptotic structure of the spacetimes by expanding the line element \eqref{metric3} in asymptotic infinity. Specifically, for the positive branch where $x\rightarrow +\infty$, we have
\bq
\label{abh}ds_+^2=-\left(1-\frac{r_S}{x}\right)dt^2+\left(1+(1+2\beta)\frac{r_S}{x}\right)dx^2+x^2d\omega^2.
\eq
Since $\beta\ll 1$, the line elements can be simplified as
\bq
\label{abh}ds_+^2=-\left(1-\frac{r_S}{x}\right)dt^2+\left(1+\frac{r_S}{x}\right)dx^2+x^2d\omega^2,
\eq
or
\bq
\label{abh}ds_+^2=-\left(1-\frac{r_S}{x}\right)dt^2+\left(1-\frac{r_S}{x}\right)^{-1}dx^2+x^2d\omega^2.
\eq
For the negative branch where $x\rightarrow-\infty$, we have the same situation after the rescaling $\tilde{x}\rightarrow-x$
\bq
\label{awh}ds_-^2=-\left(1-\frac{r_S}{\tilde{x}}\right)dt^2+\left(1-\frac{r_S}{\tilde{x}}\right)^{-1}d\tilde{x}^2+\tilde{x}^2d\omega^2.
\eq
From \eqref{abh} and \eqref{awh}, the mass of the black hole or the white hole can be read off: $M_{WH}=M_{BH}=M_0$.

In the classical limit where $x_0\rightarrow 0$, $\alpha\rightarrow 0$, and $\beta\rightarrow 0$, the line element \eqref{metric3} reduces to (the same for the $x<0$ branch):
\bq
ds_{+}^2=-\left(1-\frac{r_S}{x}\right)dt^2+\left(1-\frac{r_S}{x}\right)^{-1}dx^2+x^2d\omega^2.
\eq
Therefore, the Schwarzschild black hole solution can be recovered in the classical limit.

\section{Main properties of GOP black hole}
 In this section we would like to study the main properties of spherically symmetric loop quantum black holes, including geometric properties, energy conditions, and possible quantum-gravity effects, with particular interests at the throat, horizons, and the asymptotic infinities of the spacetime.

\subsection{The throat}
\subsubsection{Geometric property}
The most notable feature of the wormhole is that it is not smoothly connected at the throat $x=0$. In the original coordinate $x$, there is no well-defined derivative of $r$, as $\frac{dr}{dx}|_{x=+0}\neq \frac{dr}{dx}|_{x=-0}$. Figures \ref{figXx} and \ref{figrx} also show that the extension is not analytical, which implies that there exists an infinitely thin shell at the throat \cite{Wang:2010iop}. In the $(t, x, \theta, \phi)$ coordinates, we have
\bq
ds^2=-f(x)dt^2+g(x)dx^2+h^2(x)d\omega^2,
\eq
where
\bqn
\label{fx}f(x)&=&1-\frac{r_S}{|x|+x_0}+\frac{\alpha r_S^6}{(|x|+x_0)^4(|x|+x_0+r_S)^2},\\
\label{gx}g(x) &=& \left(1+\frac{r_S}{|x|+x_0}\right)\left(1+\frac{\beta r_S}{|x|+x_0}\right)^2\left[1+\frac{r_S^2((|x|+x_0)^4+(|x|+x_0)^3 r_S-\alpha r_S^4)}{(|x|+x_0)^3(|x|+x_0-r_S)(|x|+x_0+r_S)^2+\alpha r_S^6}\right],\nonumber\\
 \\
\label{hx}h(x)&=& |x|+x_0.
\eqn
By introducing the Heaviside function
\bqn
H(x) &=
\begin{cases}
1, &  x \geq 0,\cr
0, &  x <   0,\cr
\end{cases}
\eqn
one finds $|x|=H(x)x-[1-H(x)]x$, which means
\bqn
\frac{\partial |x|}{\partial x}&=&2H(x)-1 ,\\
\frac{\partial^2 |x|}{\partial x^2}&=&2\delta(x).
\eqn
We then have
 \bqn
\label{pg}
\frac{\partial G(x)}{\partial x}&=&\left(\frac{\partial G}{\partial |x|}\right)^D,\\
\frac{\partial^2 G(x)}{\partial x^2}&=&2\left(\frac{\partial G}{\partial |x|}\right)\delta(x)+\left(\frac{\partial^2 G}{\partial |x|^2}\right)^{D},\\
\label{pg1}F(x)^D &\equiv& F^+(x) H(x) + F^-(x) \left[1 - H(x)\right],
\eqn
where $G\equiv \{f, g, h\}$, and $F^{\pm}$ are functions, defined in the regions $x > 0$ and $x <0$,  respectively.
It is clear that the second-order derivatives of the functions $f(x), g(x), h(x)$ include a term proportional to $\delta(x)$. Therefore,  the energy-momentum tensor $T_{\mu\nu}$, which is defined as $T_{\mu\nu} \equiv \kappa^{-1} G_{\mu\nu}$, can be written in the following form:
\bq\label{em}
T_{\mu\nu}=H(x)T_{\mu\nu}^{+} +(1- H(x)) T_{\mu\nu}^{-} + T_{\mu\nu}^{I}\delta(x),
\eq
where $T_{\mu\nu}^{+}$ ($T_{\mu\nu}^{-}$) denotes the energy momentum for $x>0$ ($x<0$), and $T_{\mu\nu}^{I}$ represents that for an infinitely thin matter shell on the hypersurface $x = 0$.
The non-vanishing components of the Einstein tensor are
\bq
G_{\mu\nu}=H(x)G_{\mu\nu}^{+} +(1- H(x)) G_{\mu\nu}^{-}+ G_{\mu\nu}^{I} \delta(x),
\eq
with
\bqn
\label{2}
G_{tt}^{+}&=&G_{tt}^{-}=\frac{f\left(g^2+h g'h'-g\left(h'^2+2h h''\right)\right)}{h^2 g^2},\ G_{tt}^{I}=-\frac{4 f h'}{g h},\\
G_{xx}^{+}&=&G_{xx}^{-}=\frac{h f' h'+f\left(-g+h'^2\right)}{f h^2},\  G_{xx}^{I}=0\\
G_{\theta\theta}^{+}&=&G_{\theta\theta}^{-}=-\frac{h\left(f g'\Big(h f'+2f h'\right)+g\left(h\left(f'^2-2f f''\right)-2f\left(f'h'+2f h''\right)\right)\Big)}{4f^2g^2}, \\
G_{\theta\theta}^{I}&=&\frac{h\left(h f'+2f h'\right)}{f g},\\
G_{\phi\phi}^{+}&=&G_{\phi\phi}^{-}= G_{\theta\theta}^{+}\sin^2\theta=G_{\theta\theta}^{-}\sin^2\theta\label{3},\ G_{\phi\phi}^{I}=G_{\theta\theta}^{I}\sin^2\theta,
\eqn
where prime denotes differentiation with respect to $|x|$.
From \eqref{2}--\eqref{3}, we find that $G_{tt}$, $G_{\theta\theta}$, and $G_{\phi\phi}$ have a term proportional to $\delta(x)$, which can be considered as $G_{\mu\nu}^{I}$. The energy momentum is, therefore, given by $T_{\mu\nu}^{I}=\kappa^{-1} G_{\mu\nu}^{I}$.

\subsubsection{Energy conditions}
Generally speaking, the energy-momentum tensor $T_{\mu\nu}$ has the form
\begin{equation}
\label{T}
T_{\mu\nu} = \left(\rho+p\right) u_{\mu}u_{\nu} + p g_{\mu\nu}.
\end{equation}
In addition, as claimed above, one has $T_{\mu\nu}^{I}=\kappa^{-1} G_{\mu\nu}^{I}$. Therefore, together with \eqref{2}--\eqref{3},
on the throat ($x=0$) we have
\bq
\lb{EMT_Shell}
T^I_{\mu\nu} = p^I_t t_{\mu}t_{\nu} + p^I_{\theta} \left(\theta_{\mu}\theta_{\nu} + \phi_{\mu}\phi_{\nu}\right),
\eq
 where $\theta_{\mu}$ and $\phi_{\mu}$
are the unit vectors in the tangential directions of the two-sphere $t, r = $constant, and $t_{\mu}$ is the spacelike unit vector along the $dt$ direction,
as now $x$ becomes timelike near the throat, while $t$ becomes spacelike. The quantities $p^I_t$ and $ p^I_{\theta}$ denote the radial and tangential pressures of the thin shell, given,
respectively, by
\bqn
\label{rho2}
p_t^I&=&\left.\frac{4  h'}{\kappa g h}\right|_{x = 0},\\
\label{ptheta2}
p^I_{\theta}&=&\left. \frac{\left(h f'+2f h'\right)}{\kappa f g h}\right|_{x = 0}.
\eqn
The weak, the dominant, and the strong energy conditions \cite{Hawking:1973uf} are given, respectively, by\\
(i) the weak energy condition (WEC)
\begin{equation}
(i)\; \rho \geq 0, \quad (ii) \; \rho+p_x \geq 0, \quad (iii) \;\rho+p_{\theta} \geq 0,
\end{equation}
(ii) the dominant energy condition (DEC)
\begin{equation}
(i)\; \rho \geq 0,\quad (ii) \; - \rho \le  p_x \le \rho, \quad (iii) \; -\rho \le p_{\theta} \le \rho,
\end{equation}
and (iii) the strong energy condition (SEC)
\begin{equation}
 (i) \;  \rho+p_x \geq 0,\;(ii)\; \rho+p_{\theta} \geq 0,  \; (iii) \; \rho+p_x + 2p_{\theta}\geq 0.
\end{equation}

As far as the present case is concerned, $p_x$ in the above formulas should be replaced by $p_t^I$ since $x$ and $t$ exchange their roles inside the horizon, and $\rho^I = 0$.  Now taking $x=0$ into the above expressions \eqref{rho2}--\eqref{ptheta2}, we obtain
\bqn
\label{rho3}\kappa p^I_t &=&\frac{4 \left(\alpha r_S^6-r_S^3 x_0^3-r_S^2 x_0^4+r_S x_0^5+x_0^6\right)}{x_0^3 (r_S+x_0)^2 (\beta  r_S+x_0)^2},\\
\label{ptheta3}
\kappa p^I_{\theta}&=&-\frac{2 \alpha  r_S^7+4 \alpha  r_S^6x_0+r_S^4 x_0^3+r_S^3 x_0^4-3 r_S^2x_0^5-5 r_S x_0^6-2 x_0^7}{x_0^3 (r_S+x_0)^3 (\beta  r_S+x_0)^2}.
\eqn
Provided that $\delta x=\ell_{\rm Pl}$ and $\Delta=4\pi{\sqrt{3}}\gamma\ell_{\rm Pl}^2$ are chosen, one gets $\alpha=4{\sqrt{3}}\gamma\beta^2$ from \eqref{c}.\footnote{In general, the Barbero-Immirzi parameter can be taken as $\gamma\simeq 0.274$ from the ocnsiderations of black hole entropies \cite{Mei04,Agullo:2010zz,Engle:2010kt}, but we leave it as an unfixed parameter here.}  Substituting this into the above formulas \eqref{rho3}--\eqref{ptheta3} and noticing that $\beta\ll1$, we can expand the above equations at $\beta=0$  (we adopt natural units here, $\ell_{\rm Pl}=1$, $\beta=1/4M_0$),
\bqn
\label{rho4}\kappa p^I_t&=&-4\sqrt[6]{\frac{4}{3\gamma^2}}\beta^{1/3}+4\sqrt[3]{\frac{4}{3\gamma^2}}\beta^{2/3}-\frac{2\left(\sqrt{3}-12\gamma\right)}{\gamma}\beta+{\cal{O}}\left(\beta^{4/3}\right),\\
\kappa \label{ptheta4}p^I_{\theta}&=&-\sqrt[3]{\frac{9}{2\gamma^2}}\beta^{-1/3}+\frac{\sqrt{3}}{\gamma}-\frac{9-16\sqrt{3}\gamma}{2\sqrt[3]{36\gamma^4}}\beta^{1/3}+\frac{\sqrt{3}-8\gamma}{\sqrt[3]{6\gamma^5}}\beta^{2/3}+{\cal{O}}\left(\beta\right).
\eqn
From the above formulas \eqref{rho4}--\eqref{ptheta4}, it is clear that (i) $\rho^I=0$, $\rho^I+p^I_t<0$, $\rho^I+p^I_{\theta}<0$, WEC is \textit{violated}; (ii) $\rho^I=0$, $\rho^I+p^I_t<0$, $\rho^I+p^I_{\theta}<0$, $\rho^I-p^I_t>0$, $\rho^I-p^I_{\theta}>0$, DEC is \textit{violated}; (iii) $\rho^I+p^I_t<0$, $\rho^I+p^I_{\theta}<0$, $\rho^I+p^I_t+2p^I_{\theta}<0$, SEC is \textit{violated}. Therefore, none of the three energy conditions is satisfied. This confirms the expectation that the energy conditions must be violated as a result of the repulsive behavior near the transition surface, which prevents the formation of spacetime singularities.

\subsubsection{Near the throat}
The above subsections discuss the properties of the extended spacetimes exactly at the throat. In this subsection we would like to discuss the energy conditions move away but infinitely close to the throat. Equations \eqref{2}--\eqref{3} show that $G_{\mu\nu}^+=G_{\mu\nu}^-$. As a consequence it is convenient in the following sections to focus only on the $x>0$ branch. The results for the $x<0$ branch can be obtained similarly and they are the same.  In addition, in this case it turns out that using the $X$ coordinate is more suitable. It is shown in  \eqref{ax}--\eqref{X} that all derivatives of $a(X), b(X), r(X)$ with respect to $X$ are well defined except for $x=0$. As a consequence,  in terms of the $X$ coordinate, the energy density and pressure read
\bqn
\label{rho5}\kappa\rho&=&-\frac{ra_{,X}r_{,X}+a\left(-b+r_{,X}^2\right)}{abr^2},\\
\kappa p_{X}&=&-\frac{b^2+r b_{,X}r_{,X}-b\left(r_{,X}^2+2rr_{,XX}\right)}{b^2r^2},\\
\label{ptheta5}\kappa p_{\theta}&=&-\frac{a b_{,X}\left(ra_{,X}+2ar_{,X}\right)+b\left(r\left(a_{,X}^2-2aa_{,XX}\right)-2a\left(a_{,X} r_{,X}+2ar_{,XX}\right)\right)}{4a^2b^2r},
\eqn
where $_{,X}$ denotes derivative with respect to $X$.
In order to get the energy conditions near the throat, let us put $x=0$ into (\ref{X}), and we get
\begin{equation}
X_T=\frac{r_s}{x_0}.
\end{equation}
Then one can further rewrite $X_T$ through $x_0=\left(\frac{2GM_0\Delta}{4\pi}\right)^\frac{1}{3}$ and the definition of $\alpha$, as
\begin{equation}
\label{xt}
X_T=\alpha^{-\frac{1}{3}}.
\end{equation}
Now taking \eqref{xt} into above expressions \eqref{rho5}--\eqref{ptheta5}, we obtain
\bqn
\label{rho6}\kappa\rho(X_T)&=&\frac{4\pi\tilde{\alpha}\Big(3\tilde{\alpha}+5\tilde{\alpha}^2+2(1+\tilde{\alpha})^3\tilde{\alpha}\beta+(1+\tilde{\alpha})^3\beta^2\Big)}
{(1+\tilde{\alpha})^3(\tilde{\alpha}+\beta)^2\Delta},\\
\kappa p_X(X_T)&=&-\frac{4\pi\tilde{\alpha}\big((1+\tilde{\alpha})^3\beta(\beta^2+3\tilde{\alpha}\beta+2\tilde{\alpha})+\tilde{\alpha}\beta(1+3\tilde{\alpha})+\tilde{\alpha}^2(3+5\tilde{\alpha})\big)}{(1+\tilde{\alpha})^3(\tilde{\alpha}+\beta)^3\Delta},\\
\label{ptheta6}\kappa p_{\theta}(X_T)&=&\frac{2\pi\tilde{\alpha}^2\Big(\beta(9+28\tilde{\alpha}+28\tilde{\alpha}^2+8\tilde{\alpha}^3+7\tilde{\alpha}^4+2\tilde{\alpha}^5)+6\tilde{\alpha}(2+6\tilde{\alpha}+5\tilde{\alpha}^2)\Big)}
{(1+\tilde{\alpha})^4(\tilde{\alpha}+\beta)^3\Delta}.
\eqn
where $ \tilde{\alpha}=\alpha^{1/3}$.

Provided that $\delta x=\ell_{\rm Pl}$ and $\Delta=4\pi{\sqrt{3}}\gamma\ell_{\rm Pl}^2$ are chosen, one gets $\alpha=4{\sqrt{3}}\gamma\beta^2$ from \eqref{c}. Substituting this into the above formulas \eqref{rho6}--\eqref{ptheta6} and noticing that $\beta\ll1$, we can expand the above equations at $\beta=0$  (we adopt natural units here, $\ell_{\rm Pl}=1$, $\beta=1/4M$),
\bqn
\label{rho7}\kappa\rho(X_T)&=&\frac{\sqrt{3}}{\gamma}-\sqrt[3]{\frac{6}{\gamma^4}}\beta^{1/3}+\frac{3\sqrt{3}-16\gamma}{2\sqrt[3]{6\gamma^{5}}}\beta^{2/3}+\frac{10\sqrt{3}\gamma-3}{3\gamma^2}\beta+{\cal{O}}\left(\beta^{4/3}\right),\\
\kappa p_X(X_T)&=&-\frac{\sqrt{3}}{\gamma}+\sqrt[3]{\frac{6}{\gamma^4}}\beta^{1/3}-\frac{3\sqrt{3}-16\gamma}{2\sqrt[3]{6\gamma^{5}}}\beta^{2/3}+\frac{1-4\sqrt{3}\gamma}{\gamma^2}\beta+{\cal{O}}\left(\beta^{4/3}\right),\\
\kappa \label{ptheta7}p_{\theta}(X_T)&=&\frac{2\sqrt{3}}{\gamma}-\frac{9}{2}\sqrt[3]{\frac{3}{4\gamma^4}}\beta^{1/3}-\frac{48\gamma-15\sqrt{3}}{4\sqrt[3]{6\gamma^5}}\beta^{2/3}+\frac{56\sqrt{3}\gamma-33}{12\gamma^2}\beta+{\cal{O}}\left(\beta^{4/3}\right).
\eqn
From the above formulas \eqref{rho7}--\eqref{ptheta7}, it is direct to show that (i) $\rho>0$, $\rho+p_X<0$, $\rho+p_{\theta}>0$, WEC is \textit{violated}; (ii) $\rho>0$, $\rho+p_X<0$, $\rho+p_{\theta}>0$, $\rho-p_X>0$,  $\rho-p_{\theta}<0$, DEC is \textit{violated}; (iii) $\rho+p_X<0$, $\rho+p_{\theta}>0$, $\rho+p_X+2p_{\theta}>0$, SEC is \textit{violated}. Therefore, none of the three energy conditions are satisfied.
\subsubsection{Ricci and Kretschmann scalars}
To understand the geometry underlying the effective metric \eqref{metric1}, it is useful to examine the Ricci and Kretschmann scalars of the GOP quantum black hole,
\bqn
R&=&\frac1{2a^2b^2r^2}\times\big[br^2a_{,X}^2+ar(-4ba_{,X}r_{,X}+r(a_{,X}b_{,X}-2ba_{,XX}))+4a^2(b^2+r b_{,X}r_{,X}\nb\\
&&-b(r_{,X}^2+2rr_{,XX}))\big],\\
{\cal{K}}&=&R_{\mu\nu\rho\sigma} R^{\mu\nu\rho\sigma}=\frac1{4a^4b^4r^4}\times\big[b^2r^4a_{,X}^4+a^2(8b^2r^2a_{,X}^2r_{,X}^2+r^4(a_{,X}b_{,X}-2ba_{,XX})^2)\nb\\
&&+2abr^4a_{,X}^2(a_{,X}b_{,X}-2ba_{,XX})+8a^4(2b^4-4b^3r_{,X}^2+r^2b_{,X}^2r_{,X}^2-4br^2b_{,X}r_{,X}r_{,XX}\nb\\
&&+2b^2(r_{,X}^4+2r^2r_{,XX}^2))\big].
\eqn
At the throat, the expressions of the Ricci scalar and Kretschmann scalar are
\bqn
\label{RS}R &=&-\frac{24 \pi}{\Delta}+{\cal{O}}\left(\beta^{1/3}\right),\\
\label{KS}{\cal{K}}&=&\frac{5760\pi^2}{\Delta^2}+{\cal{O}}\left(\beta^{1/3}\right).
\eqn
These upper bounds agree with the results obtained in the GOP polymer black hole \cite{Gambini:2020nsf}. We know that $R$ and $\cal{K}$ diverge at the singularity in GR, but now it can be clearly seen that $R\sim\ell_{\rm Pl}^{-2}$ and ${\cal{K}}\sim\ell_{\rm Pl}^{-4}$ deviate from the classical case due to the quantum effects.
\subsection{The horizon}
Now let us turn to the horizon of the spacetime which corresponds to $a(X)=0$. Since $a(X)$ includes a sixth-order polynomial in $X$, we cannot obtain its analytic solution. Instead, one can solve the algebraic equation by iterations. From the expression for $a(X)$, we can clearly see that the root of $a(X) =0$ is close to $X=1$,  so we construct an expression for $X=1+\frac{\alpha X^6}{(1+X)^2}$, and insert $X=1$ into the right-hand side of the expression. After one iteration we have $X=1+\frac{\alpha}{4}$. In order to be more accurate, we carry out a second iteration by putting the results back to the right-hand side. Since $\alpha$ is of the order $\ell_{\rm Pl}^2$, it is precise enough to truncate it at the order $\alpha^2$, then we have
\bq
X_H=1+\frac{\alpha}{4}+\frac{5\alpha^2}{16},
\eq
which corresponds to
\bq
\label{xh}
x_H=r_S\left(1-\alpha^{1/3}-\frac{\alpha}{4}-\frac{\alpha^2}{4}\right).
\eq
In the white hole regime, it is similar to obtain
\bq\label{xh'}
x_{H'}=-r_S\left(1-\alpha^{1/3}-\frac{\alpha}{4}-\frac{\alpha^2}{4}\right).
\eq

\subsubsection{Energy conditions}
At the horizon, it turns out that one cannot judge whether the energy conditions are violated or not until we expand it up to the order of  $\beta^9$. Substituting $X_H$ into the expressions for energy density and pressure
\bqn
\label{rho8}\kappa\rho&=&\frac{b^2+r b_{,X}r_{,X}-b\left(r_{,X}^2+2rr_{,XX}\right)}{b^2r^2},\\
\kappa p_{X}&=&\frac{ra_{,X}r_{,X}+a\left(-b+r_{,X}^2\right)}{abr^2},\\
\label{ptheta8}\kappa p_{\theta}&=&-\frac{a b_{,X}\left(ra_{,X}+2ar_{,X}\right)+b\left(r\left(a_{,X}^2-2aa_{,XX}\right)-2a\left(a_{,X}r_{,X}+2ar_{,XX}\right)\right)}{4a^2b^2r}.
\eqn
Then we obtain
\bqn
\kappa\rho(X_H)&=&8 \beta ^3+4  \left(4 \sqrt{3} \gamma -3\right)\beta ^4+8\left(2- \sqrt{3} \gamma \right)\beta ^5+10\left(33 \gamma ^2-2\right)\beta ^6\nb\\
&&-4  \left(81 \gamma ^2-4 \sqrt{3} \gamma -6\right)\beta ^7 + \left(2607\sqrt{3}\gamma^3+342\gamma^2-40\sqrt{3}\gamma-28\right)\beta ^8\nb\\
&&- 4\left(1611\sqrt{3}\gamma^3+54\gamma^2-18\sqrt{3}\gamma-8\right)\beta ^9+{\cal{O}}\big(\beta^{10}\big),\\
\kappa p_X(X_H)&=&-8 \beta ^3+4\left(3-4 \sqrt{3} \gamma \right)\beta ^4 +8  \left(\sqrt{3} \gamma -2\right)\beta ^5+ 10\left(2-33 \gamma ^2\right)\beta ^6\nb\\
&&+4 \left(81 \gamma ^2-4 \sqrt{3} \gamma -6\right)\beta ^7 -\left(2607\sqrt{3}\gamma^3+342\gamma^2-40\sqrt{3}\gamma-28\right)\beta ^8\nb\\
&&+2 \left(2805\sqrt{3}\gamma^3+108\gamma^2-36\sqrt{3}\gamma-16\right)\beta ^9++{\cal{O}}\big(\beta^{10}\big),\\
\kappa p_{\theta}(X_H)&=&2\beta^3+3\left(13\sqrt{3}\gamma-2\right)\beta^4+4\left(3-20\sqrt{3}\gamma\right)\beta^5+\left(782\gamma^2+117\sqrt{3}\gamma-20\right)\beta^6\nb\\
&&-3\left(633\gamma^2+48\sqrt{3}\gamma-10\right)\beta^7+\left(12429\sqrt{3}\gamma^3/2+3249\gamma^2+155\sqrt{3}\gamma-42\right)\beta^8\nb\\
&&-\left(16377\sqrt{3}\gamma^3+4737\gamma^2+144\sqrt{3}\gamma-56\right)\beta^9++{\cal{O}}\big(\beta^{10}\big).
\eqn
 The results clearly show that $\rho+p_X<0$, which means that the three energy conditions are violated at the horizon.
\subsubsection{Ricci and the Kretschmann scalars }
To study the effects of quantum gravity further, we again would like to learn the Ricci scalar and the Kretschmann scalar near the horizon,
\bqn
R&=&\frac{48\sqrt{3}\pi\beta^3 \gamma}{\Delta}+{\cal{O}}\left(\beta^4\right),\\
\label{hk}{\cal{K}}&=&\frac{9216 \pi ^2 \beta ^4 \gamma ^2}{\Delta ^2}+{\cal{O}}\left(\beta^5\right).
\eqn
Comparing with \eqref{RS} and \eqref{KS}, we see that the Ricci scalar decreases and switches sign as one moves toward the horizon, the Kretschmann scalar also decreases fast away from the most quantum region. This is the result of quantum corrections in these effective geometries.
But for a low-mass black hole, the curvature can also be large near the event horizon.

\subsection{The spatial infinities}
\subsubsection{Energy conditions}
Now we turn our attention toward spatial infinities $x\rightarrow\pm\infty$. At the two asymptotically flat regions, we find that
\bqn
\kappa\rho(x) &\approx&
\begin{cases}
\frac{8M_0+3}{4x^4}+{\cal{O}}\left(\epsilon^5\right), & x \rightarrow \infty,\cr
\frac{8M_0+3}{4x^4}+{\cal{O}}\left(\epsilon^5\right), &  x \rightarrow - \infty,\cr
\end{cases}\nb\\
\kappa p_X(x) &\approx&
\begin{cases}
-\frac1{x^3}+\frac{3}{4x^4}+{\cal{O}}\left(\epsilon^5\right), &  x \rightarrow \infty,\cr
\frac1{x^3}+\frac{3}{4x^4}+{\cal{O}}\left(\epsilon^5\right), &  x \rightarrow - \infty,\cr
\end{cases}\nb\\
\kappa p_{\theta}(x) &\approx&
\begin{cases}
\frac1{2x^3}-\frac{2M_0+3}{4x^4}+{\cal{O}}\left(\epsilon^5\right), &  x \rightarrow \infty,\cr
-\frac1{2x^3}-\frac{2M_0+3}{4x^4}+{\cal{O}}\left(\epsilon^5\right), &  x \rightarrow - \infty,\cr
\end{cases}\nb\\
\eqn
where $\epsilon \equiv 1/|x|$.
Thus, none of the three energy conditions holds in these two asymptotically flat regions.
\subsubsection{Ricci and Kretschmann scalars}
Likewise, we want to know what happens to the Ricci scalar and the Kretschmann scalar at asymptotically flat regions,
\bqn
R &\approx&
\begin{cases}
\frac{3\left(1+2M_0\right)}{2x^4}-\frac{\left(9+4M_0\right)}{2x^5}+{\cal{O}}\left(\epsilon^6\right), &  x \rightarrow \infty,\cr
\frac{3\left(1+2M_0\right)}{2x^4}+\frac{\left(9+4M_0\right)}{2x^5}+{\cal{O}}\left(\epsilon^6\right)&  x \rightarrow - \infty,\cr
\end{cases}\nb\\
{\cal{K}} &\approx&
\begin{cases}
\frac{6\left(1+4M_0+8M_0^2\right)}{x^6}+{\cal{O}}\left(\epsilon^7\right), &  x \rightarrow \infty,\cr
\frac{6\left(1+4M_0+8M_0^2\right)}{x^6}+{\cal{O}}\left(\epsilon^7\right). &  x \rightarrow - \infty.\cr
\end{cases}\nb\\
\eqn
From these expressions, we observe that both the Ricci scalar and the Kretschmann scalar have the same behavior as the classical Schwarzschild solution at the spatial infinities.

\begin{figure}[h!]
{\centering
  \includegraphics[width = 0.5\textwidth]{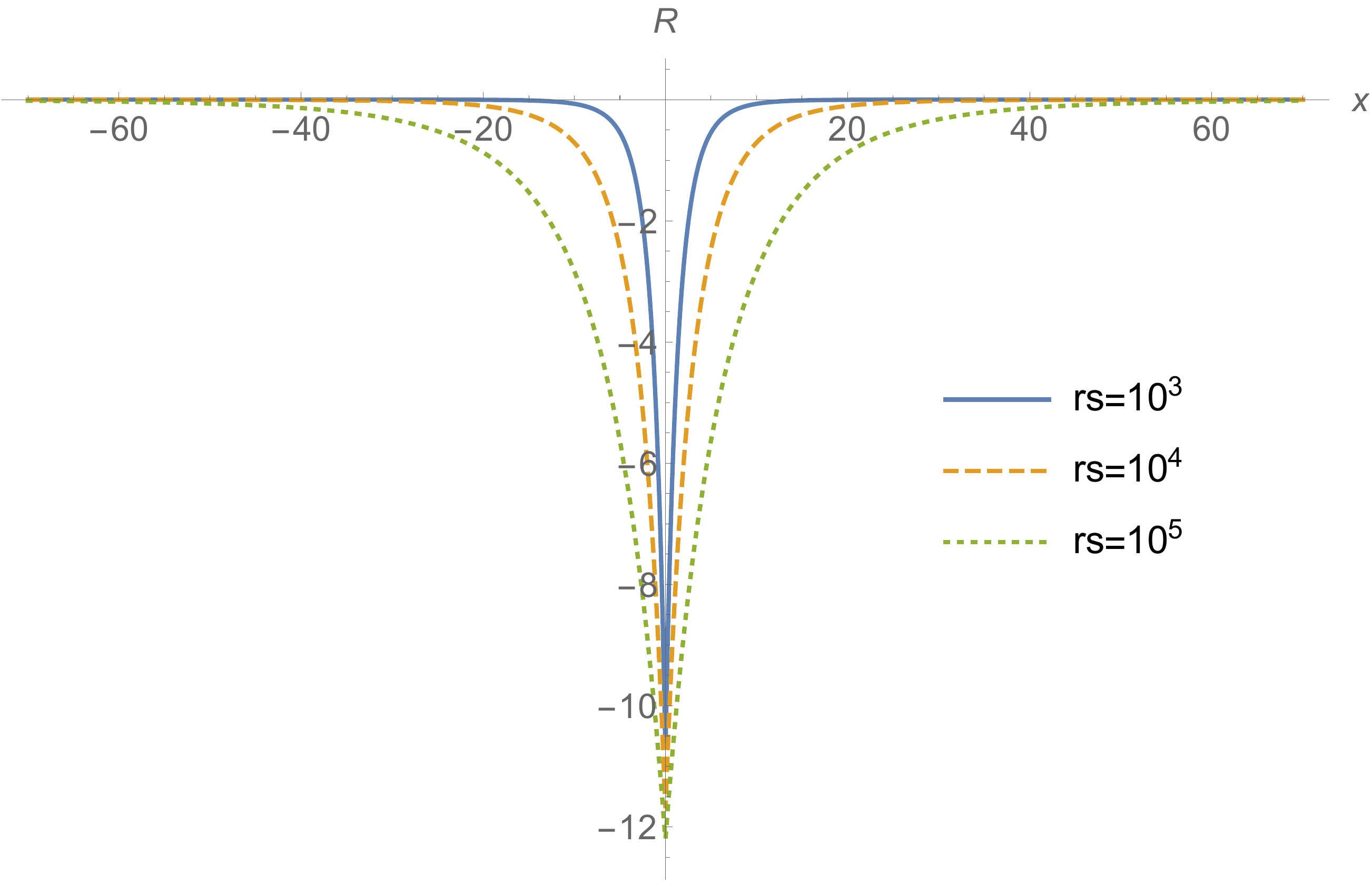}
}
\caption{Ricci scalar $R$ for different value of the mass $M_0$. The solid, dashed, and dotted curves correspond to $M_0=10^3$, $M_0=10^4$, and $M_0=10^5$ respectively. Here we set $\ell_{\rm Pl}=1$.}
\label{figR}
\end{figure}

\begin{figure}[h!]
{\centering
  \includegraphics[width = 0.5\textwidth]{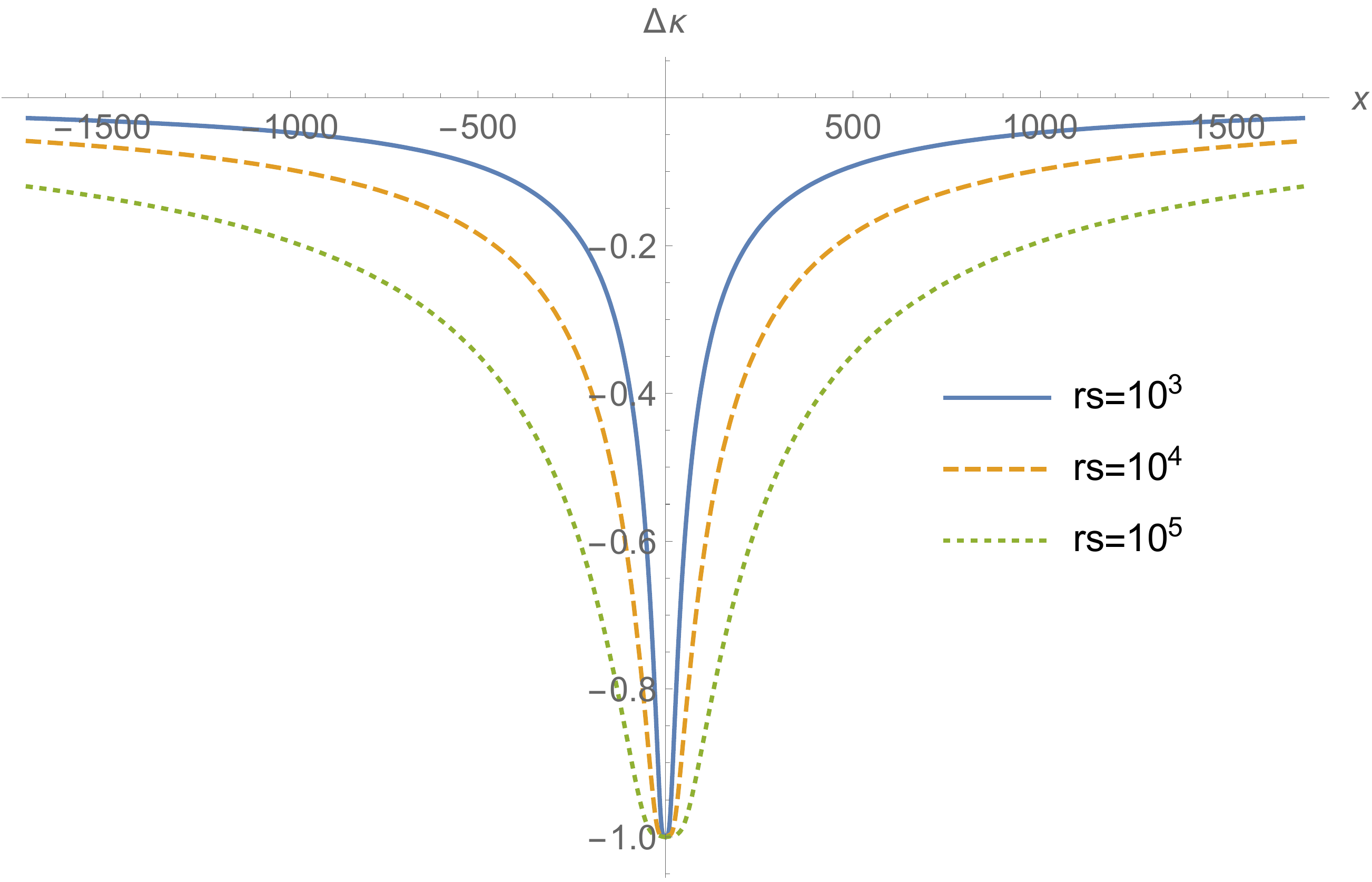}
}
\caption{Relative difference of the Kretschmann scalar for different value of the mass $M_0$. The solid, dashed, and dotted curves correspond to $M_0=10^3$, $M_0=10^4$, and $M_0=10^5$ respectively. Here we set $\ell_{\rm Pl}=1$. }
\label{figK}
\end{figure}

In order to investigate the deviation from GR in more details, let us consider the Ricci scalar and the relative difference $\Delta \cal{K}$ of the Kretschmann scalar for the whole spacetime, and $\Delta \cal{K}$ is defined by
\bq
\Delta {\cal{K}} \equiv \frac{{ {\cal{K}}-\cal{K}}^{GR} }{{\cal{K}}^{GR}},
\eq
where ${\cal{K}}^{GR}$ denotes the Kretschmann scalar of the Schwarzschild solution. In GR, we have $R^{GR}=0$ except at singularity, and ${\cal{K}}^{GR}$ is given by
\bqn
{\cal{K}}^{GR} &\equiv& R_{\alpha\beta\mu\nu}R^{\alpha\beta\mu\nu} =
 \begin{cases}
\frac{48 M_{BH}^2}{r^6(x)}, & x > 0,\cr
\frac{48 M_{WH}^2}{r^6(x)}. & x < 0.\cr
\end{cases} ~~~~~
\eqn
Therefore, we show the deviations of the quantum black holes from the classical Schwarzschild black hole in Figs. \ref{figR} and \ref{figK}.

From Figs. \ref{figR} and \ref{figK} we can clearly see that the quantum effect is significant near the throat, but it becomes negligible once it leaves the throat, and eventually approaches to the classical situation at asymptotically flat regions.
\section{Quasinormal modes}
In this section, we are going to study the perturbations of the GOP quantum black hole and their associated QNMs. Technically, there are various methods to calculate the QNM frequencies, ranging from numerical approaches \cite{Leaver:1986gd, Jansen:2017oag} to semianalytic methods \cite{Nollert:1999ji, Berti:2009kk, Konoplya:2011qq}. Among these technical methods, we will use a semianalytical approach, which is constructed on the Wentzel-Kramers-Brillouin (WKB) approximation, to evaluate the QNM frequencies. In the rest of this section, on the one hand, we discuss the perturbations of massless scalar fields, electromagnetic fields, and axial gravitational fields. On the other hand, we calculate the QNM frequencies in the Eikonal limit.

\subsection{The massless scalar field perturbations}
\begin{figure}[h!]
{\centering
  \includegraphics[width = 0.7\textwidth]{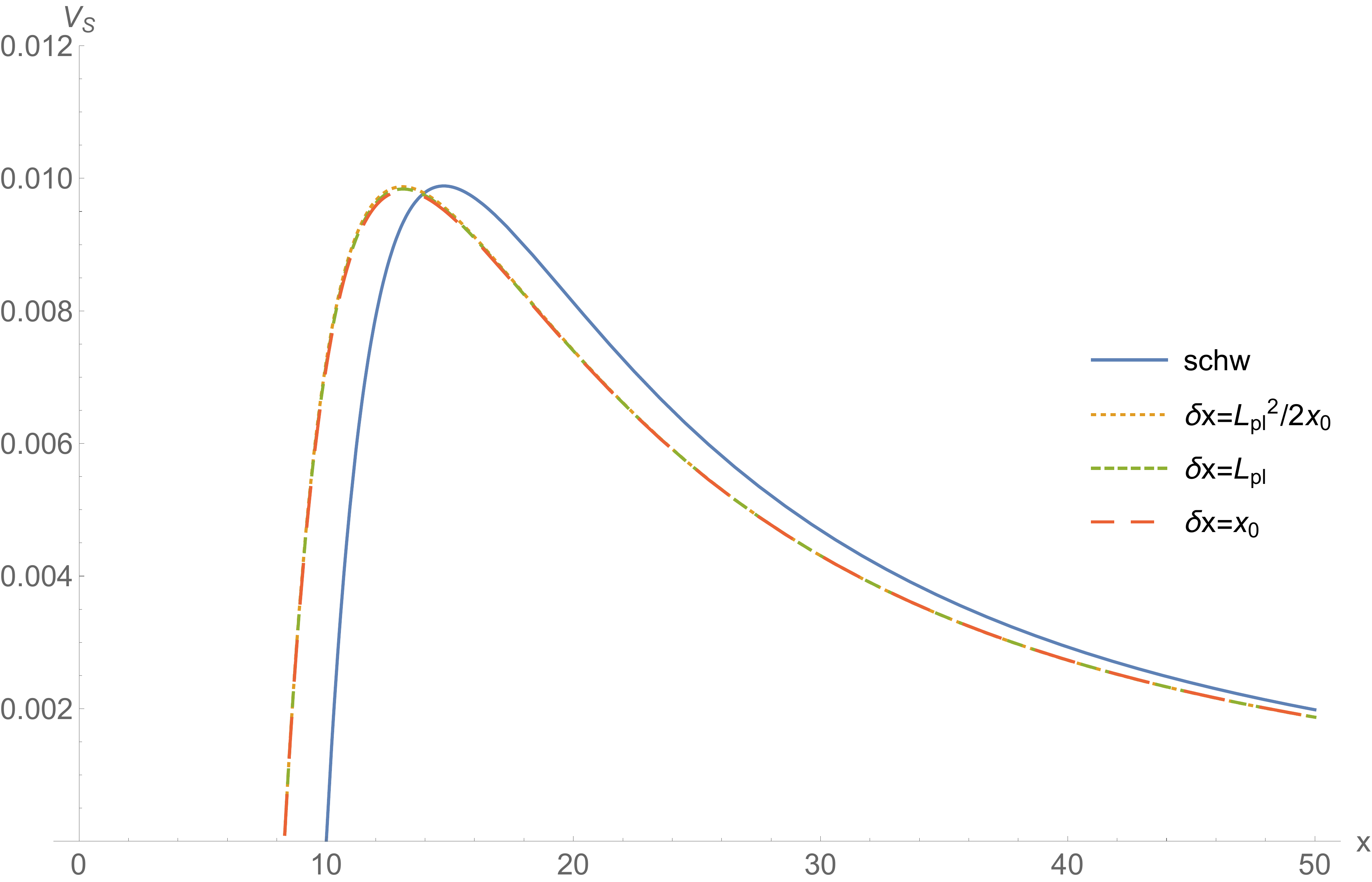}
}
\caption{The effective potential $V_s(x)$ is shown for different values of parameters. We know that $\delta x \in [\ell_{\rm Pl}^2/2 x_0,x_0]$, the dotted, short dash, and long dash curves correspond to $\delta x=\ell_{\rm Pl}^2/2x_0$, $\delta x=\ell_{\rm Pl}$, and $\delta x=x_0$ respectively. The potential corresponding to the Schwarzschild solution is presented by the solid curves. Here we applied $M_0=10$ and the multipole number $l=2$. }
\label{figVs}
\end{figure}

We first consider the simplest example, a massless scalar field in the background of a black hole spacetime, which obeys the Klein-Gordon equation
\bq
\label{Phi} \square \Phi=0,
\eq
where $\square=\bigtriangledown^\mu\bigtriangledown_\mu$. When the spacetime is spherically symmetric, the line element can be cast as
\bq
ds^2=-|g_{tt}|dt^2+g_{rr}dr^2+r^2d\omega^2.
\eq
Comparing with the metric \eqref{metric3}, we can see that $r$ is the same as $x+x_0$. For this case, the solution to \eqref{Phi} can always be decomposed into spherical harmonics $Y_{\ell m}$ as
\bq
\Phi=\sum_{\ell m} \frac{\psi_{\ell m}(r)}{r} Y^{\ell m}(\theta,\phi) e^{-i\omega t},
\eq
where $\ell$ and $m$ are the spherical harmonic indices. The radial part of the scalar field satisfies the following equation:
\bq
\partial_{r_{\ast}}^2 \psi_{\ell}+Q(r_{\ast})\psi_{\ell}=0,
\label{scalarr}
\eq
where $Q(r_{\ast})=\omega^2-V_S(r)$, and $r_{\ast}$ is the tortoise coordinate defined as
\bq
\frac{d r_{\ast}}{d r}=\sqrt{\frac{g_{rr}}{|g_{tt}|}}.
\label{tort}
\eq
The effective potential $V_S$ is given by
\bq
V_s(r)=|g_{tt}|\left[\frac{l(l+1)}{r^2}+\frac1{r \sqrt{|g_{tt}|g_{rr}}}\left(\frac{d}{d r}\sqrt{\frac{|g_{tt}|}{g_{rr}}}\right)\right].
\eq
It should be noticed that in the above derivations, we have taken $r=x+x_0$ as the radial coordinate. In the $x$ coordinate however, we have
\bq
\frac{d r_{\ast}}{d x}=\sqrt{\frac{g_{xx}}{|g_{tt}|}},
\label{tortx}
\eq
\bq
V_s(x)=|g_{tt}|\left[\frac{l(l+1)}{(x+x_0)^2}+\frac1{(x+x_0)\sqrt{|g_{tt}|g_{xx}}}\left(\frac{d}{d x}\sqrt{\frac{|g_{tt}|}{g_{xx}}}\right)\right],
\eq
with
\bqn
\label{gt}g_{tt}&=&\left(1-\frac{r_S}{x+x_0}+\frac{\alpha r_S^6}{(x+x_0)^4(x+x_0+r_S)^2}\right),\\
\label{gx}g_{xx}&=&\left(1+\frac{r_S}{x+x_0}\right)\left(1+\frac{\beta r_S}{x+x_0}\right)^2\left[1+\frac{r_S^2((x+x_0)^4+(x+x_0)^3 r_S-\alpha r_S^4)}{(x+x_0)^3(x+x_0-r_S)(x+x_0+r_S)^2+\alpha r_S^6}\right].
\eqn
In the case of a Planckian black hole, the difference of the effective potential between the Schwarzschild black hole and the quantum black hole becomes apparent in the massless scalar field perturbation as shown in Fig. \ref{figVs}.
Then we are going to use the WKB approximation method to calculate QNM frequencies. We impose the boundary conditions, that is, there are only ingoing waves moving toward the black hole at the horizon which means that nothing can escape from the event horizon, and there are only outgoing waves moving away from the black hole at spatial infinity.
From (\ref{scalarr}), we find that the form of the potential is the same as that of a $Schr\ddot{o}dinger$ equation with a potential barrier $Q(r_{\ast})$, and the QNM frequencies depend on the behavior of the potential \cite{Iyer:1986np} as the following:
\bq
\label{Q0}
\frac{i Q_0}{\sqrt{2 Q_0''}}-\sum_{i=2}^{6}\Lambda_i=n+\frac{1}{2},
\eq
where $Q_0=\omega^2-V(x)|_{peak}$, a prime indicates differentiation with respect to $r_{\ast}$, and $n$, $\Lambda_i$ are, respectively, the overtone number and the WKB correction terms. The expressions for $\Lambda_{2,3}$ were derived in \cite{Iyer:1986np} and $\Lambda_{4,5,6}$ were derived in \cite{Konoplya:2003ii}. As to the present work, it turns out that the third-order approximation is precise enough. Therefore in what follows we compute the QNM frequencies only up to the third-order correction.
\begin{figure}[htb]
{\centering
  \includegraphics[width = 0.7\textwidth]{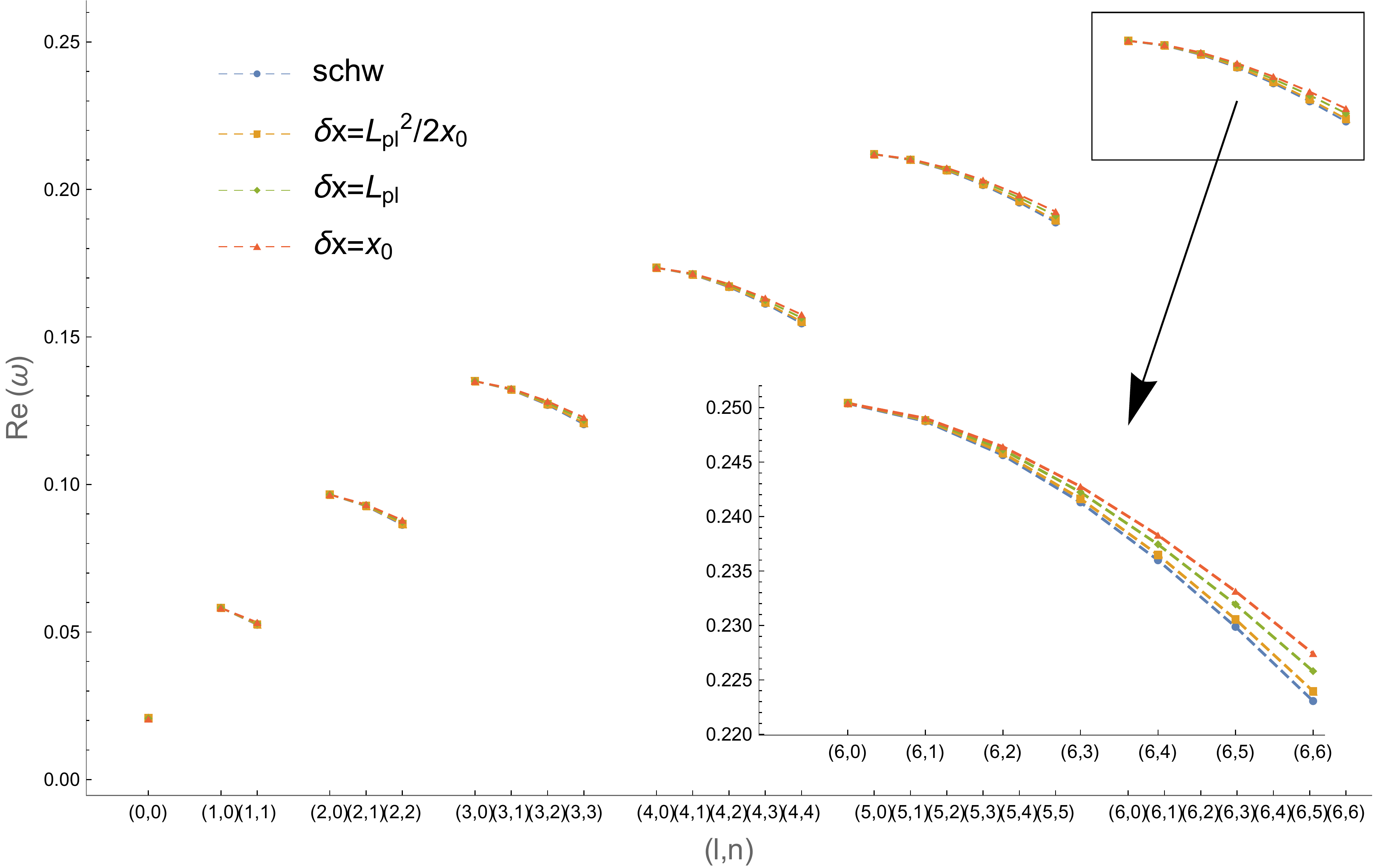}\hspace{0.30cm}
  \includegraphics[width = 0.7\textwidth]{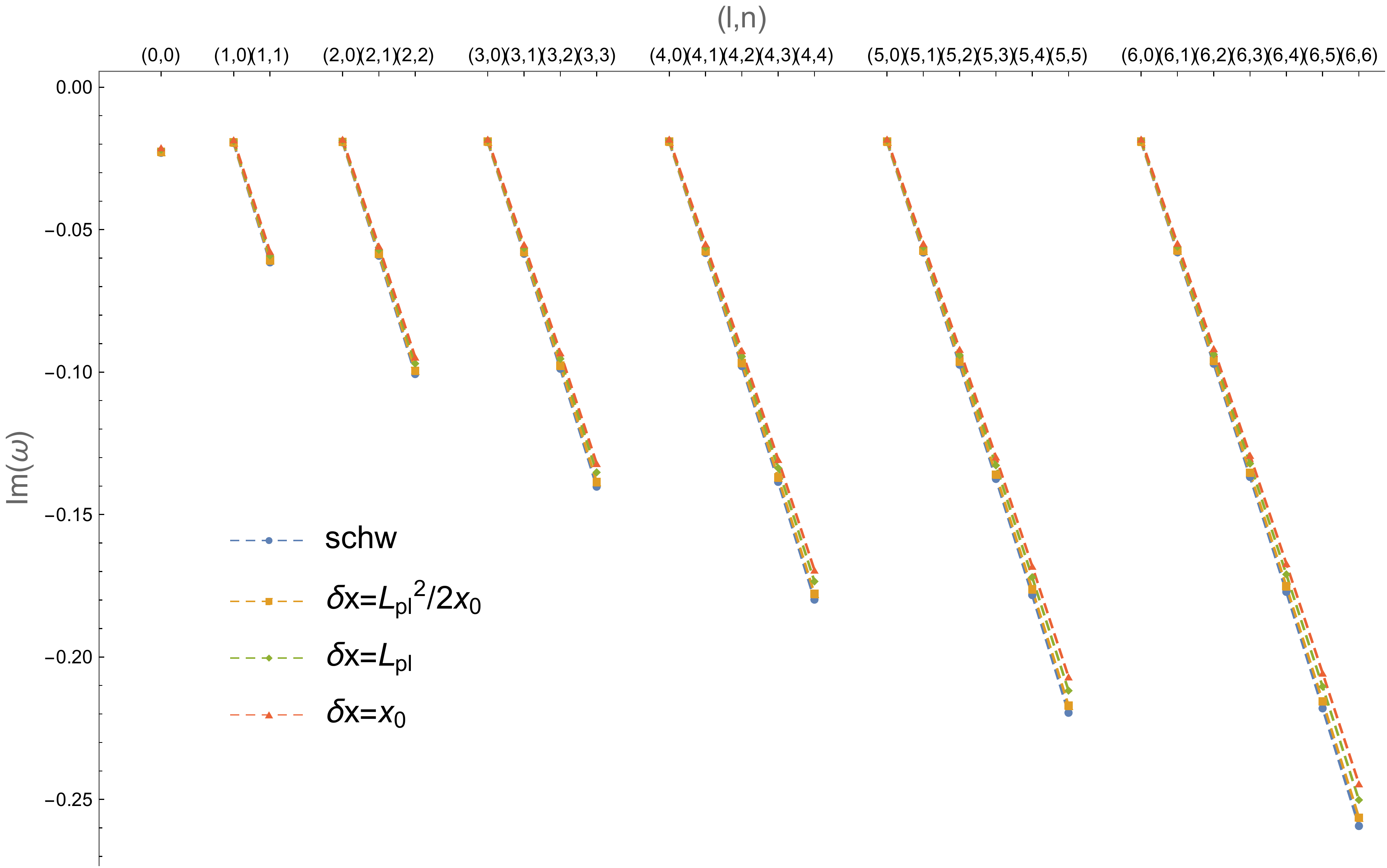}
}
\caption{The real part (top) and the imaginary part (bottom) of the QNMs for the massless scalar field perturbation are presented. Different markers represent different values of parameters $\delta x$: the square, diamond, and triangle correspond to $\delta x=\ell_{\rm Pl}^2/2x_0$, $\delta x=\ell_{\rm Pl}$, and $\delta x=x_0$ respectively. The Schwarzschild case is presented by the point.  The panels (top and bottom) show how frequencies change with respect to the change of the multipole number $l$ and overtone number $n$. The data used here are shown in Table \ref{tbs}. }
\label{figsw}
\end{figure}

For comparison, we provide the QNM frequencies for the multipole number $l=0, 1, 2, 3, 4, 5$ and $6$ for different $\delta x$ in Table \ref{tbs} (as can be seen in the Appendix), and plot QNM frequencies of the quantum black hole for different values of the parameter $\delta x$ in Fig. \ref{figsw}.

From Fig. \ref{figsw}, one can see that the real part of the quantum black hole frequency $\omega_R$ (the oscillation frequency of perturbations) is larger than that of the Schwarzschild black hole except for $n=0$. As the quantum parameter $\delta x$ moves away from the classical case, the real part of the frequency becomes higher and is less sensitive to $n$. As for the damping rate ($|\omega_I|$) of the quantum black hole, which is lower than that of the Schwarzschild black hole, it becomes slower when the quantum parameter moves away from the classical one. This trend is the same as the result of the Ashtekar-Olmedo-Singh (AOS) quantum black holes \cite{Ashtekar:2020ckv,Daghigh:2020fmw} but is different from the Bodendorfer-Mele-M\"unch (BMM) black hole \cite{Bouhmadi-Lopez:2020oia}.

\subsection{The electromagnetic perturbations}
\begin{figure}[h!]
{\centering
  \includegraphics[width = 0.7\textwidth]{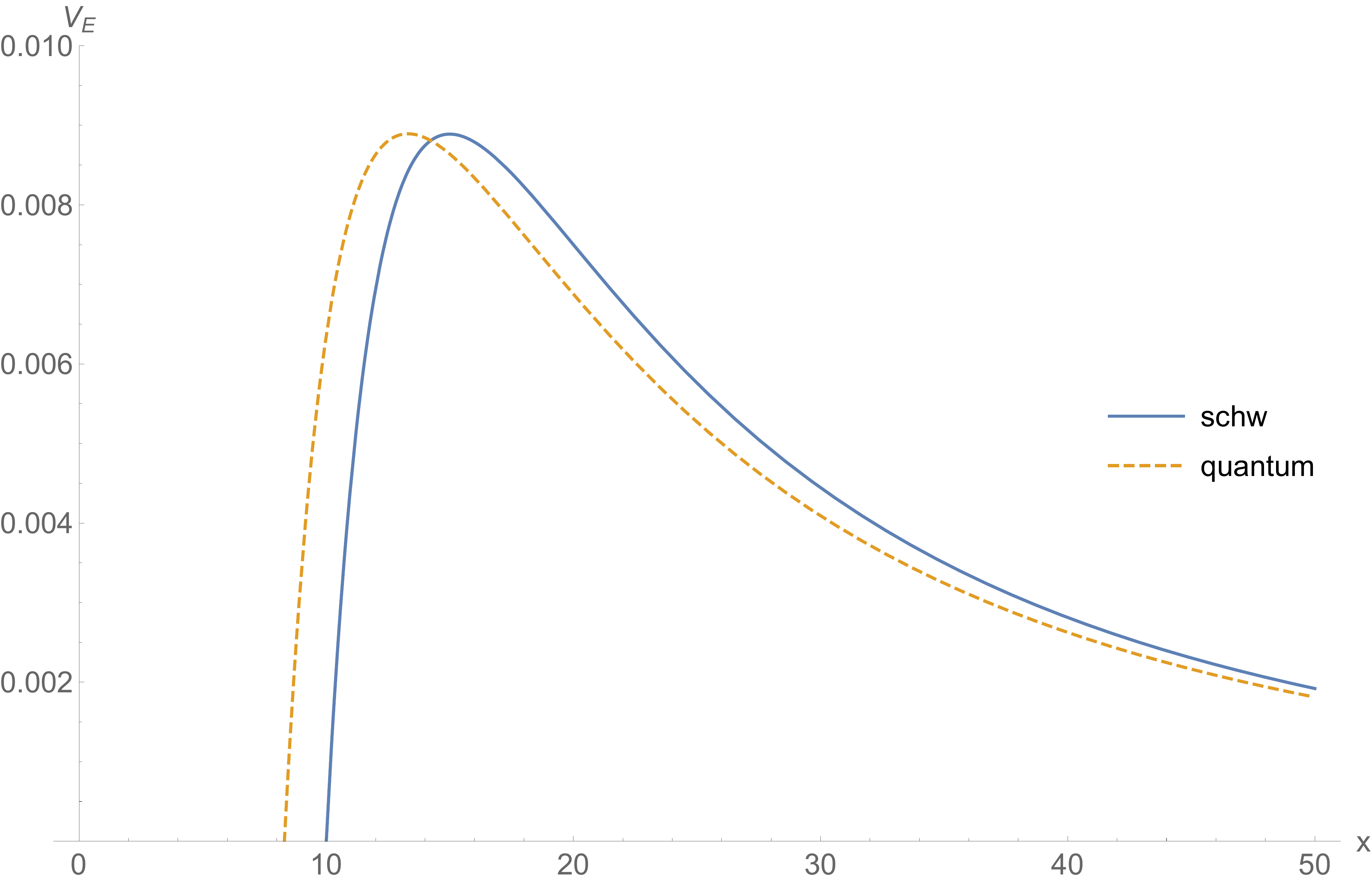}
}
\caption{The effective potential $V_E(x)$. The dashed and solid curves correspond to the quantum black hole and Schwarzschild solution. Here we applied $M_0=10$ and the multipole number $l=2$}
\label{figVe}
\end{figure}

In this subsection, we discuss QNM frequencies of electromagnetic perturbations around the GOP quantum black holes, whose master equation is the Maxwell equation. In the tetrad formalism \cite{Chandrasekhar:1985kt}, we have the Bianchi identity of the field strength $F_{[(a)(b)|(c)]}=0$ and the conservation equation $\eta^{(n)(m)}(F_{(a)(n)})_{|(m)}=0$, which give
\bqn
&(r \sin\theta \sqrt{g_{rr}} F_{(\phi)(r)})_{,\theta}+(r^2 \sin\theta F_{(\theta)(\phi)})_{,r}=0,\label{e1}\\
&(r \sqrt{|g_{tt}|}F_{(t)(\phi)})_{,r}+r \sqrt{g_{rr}} F_{(\phi)(r),t}=0,\\
&r \sqrt{|g_{tt}|} (F_{(t)(\phi)} \sin\theta)_{,\theta}+r^2 \sin\theta F_{(\phi)(\theta),t},=0,\label{e2}
\eqn
and
\bq
\label{e3}
(r \sqrt{|g_{tt}|} F_{(\phi)(r)})_{,r}+\sqrt{|g_{tt}| g_{rr}}F_{(\phi)(\theta),\theta}+r \sqrt{g_{rr}} F_{(t)(\phi),t}=0.
\eq
Differentiating \eqref{e3} with respect to $t$, and replacing $F_{(\phi)(r)}$, $F_{(\phi)(\theta)}$ with $F_{(t)(\phi)}$ by \eqref{e1}--\eqref{e2}, we find
\bq
\Big[\sqrt{\frac{|g_{tt}|}{g_{rr}}}(r\sqrt{|g_{tt}|}\mathcal{B})_{,r}\Big]_{,r}+\frac{|g_{tt}|\sqrt{g_{rr}}}{r}\Big(\frac{\mathcal{B}_{,\theta}}{\sin\theta}\Big)_{,\theta}\sin\theta-r\sqrt{g_{rr}}\mathcal{B}_{,tt}=0,
\label{e4}
\eq
with
\bq
\mathcal{B}\equiv F_{(t)(\phi)}\sin\theta.
\eq
Make the following field decomposition
\bq
\label{e5}
\mathcal{B}=\sum_{\ell m} \varphi_{\ell m}(r) C^{-1/2}_{l+1}(\theta) e^{-i\omega t},
\eq
where $C^{v}_{n}$ is the Gegenbauer function \cite{Abramowitz:1980mi} governed by the equation
\bq
\big[\frac{d}{d\theta} \sin^{2v}\theta\frac{d}{d\theta}+n(n+2v)\sin^{2v}\theta\big]C^{v}_{n}(\theta)=0.
\eq
It should be noted that the Gegenbauer function $C^{-1/2}_{l+1}(\theta)$ can be read
\bq
\sin\theta \frac{d}{d\theta}(\frac1{\sin\theta}\frac{d}{d\theta}C(\theta))=-l(l+1)C(\theta).
\eq
With the substitution \eqref{e5}, and Eq. \eqref{e4} can be transformed to
\bq
\label{elect} \Big[\sqrt{\frac{|g_{tt}|}{g_{rr}}}(r\sqrt{|g_{tt}|}\varphi(r))_{,r}\Big]_{,r}+\omega^2r\sqrt{g_{rr}} \varphi(r)-\frac{|g_{tt}|\sqrt{g_{rr}}}{r}l(l+1) \varphi(r)=0.
\eq
Then we redefine $\psi_{EM}\equiv r\sqrt{|g_{tt}|} \varphi(r)$, bring $x$ back, and use the tortoise radius defined by (\ref{tortx}). Equation (\ref{elect}) can be written as
\bq
\partial_{r_{\ast}}^2 \psi_{EM}+(\omega^2-V_E(x))\psi_{EM}=0,
\eq
and the effective potential of electromagnetic perturbation $V_E(x)$ have the following form \cite{Chandrasekhar:1985kt}:
\bq
V_E(x)=|g_{tt}|\frac{l(l+1)}{(x+x_0)^2},
\eq
where the $g_{tt}$ is the same as in \eqref{gt}. We plot the the effective potential $V_E(x)$ in Fig. \ref{figVe}, where we see that the black hole with the electromagnetic field perturbation is different from the massless scalar field perturbation.

\begin{figure}[htb]
{\centering
  \includegraphics[width = 0.7\textwidth]{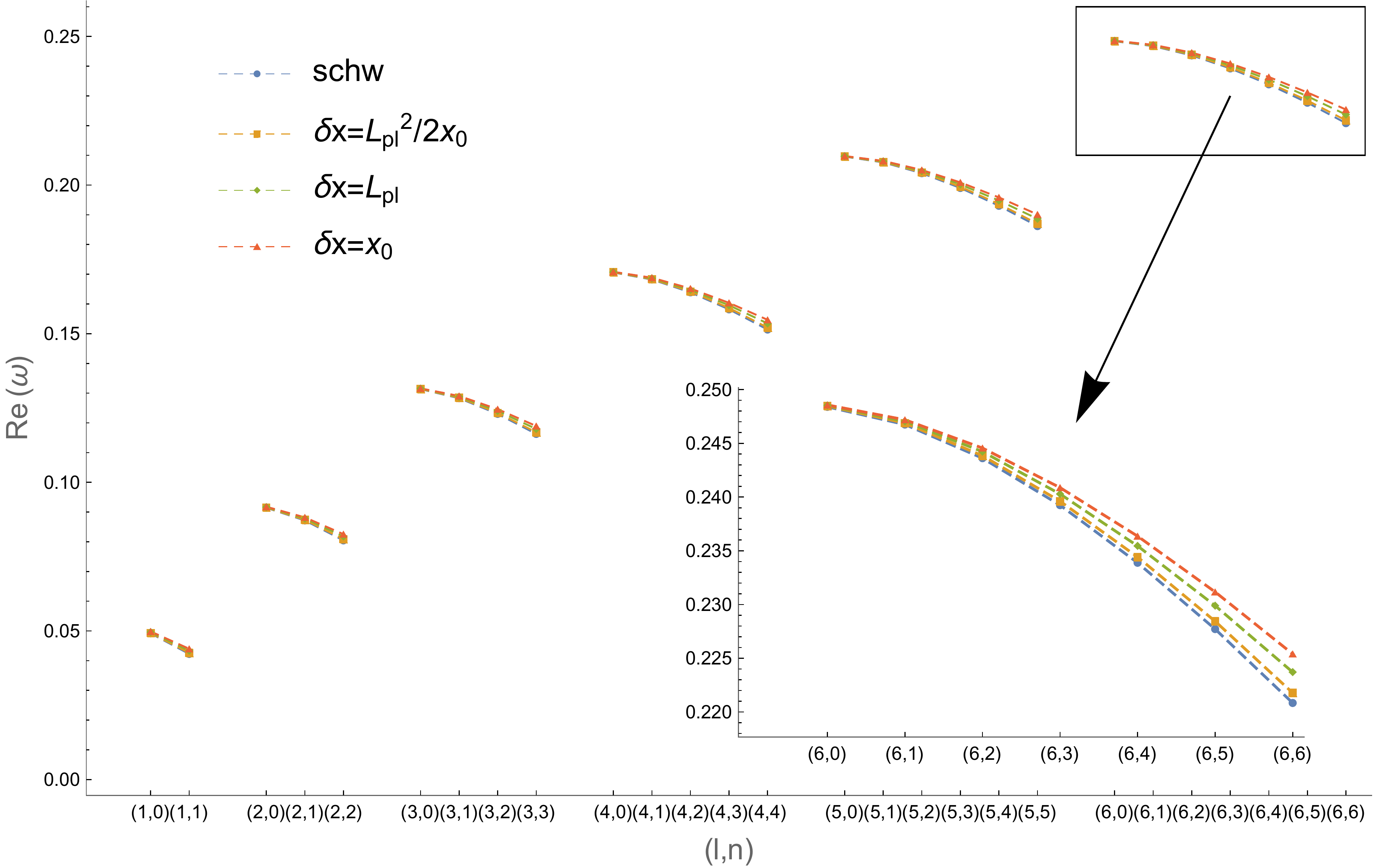}\hspace{0.45cm}
  \includegraphics[width = 0.7\textwidth]{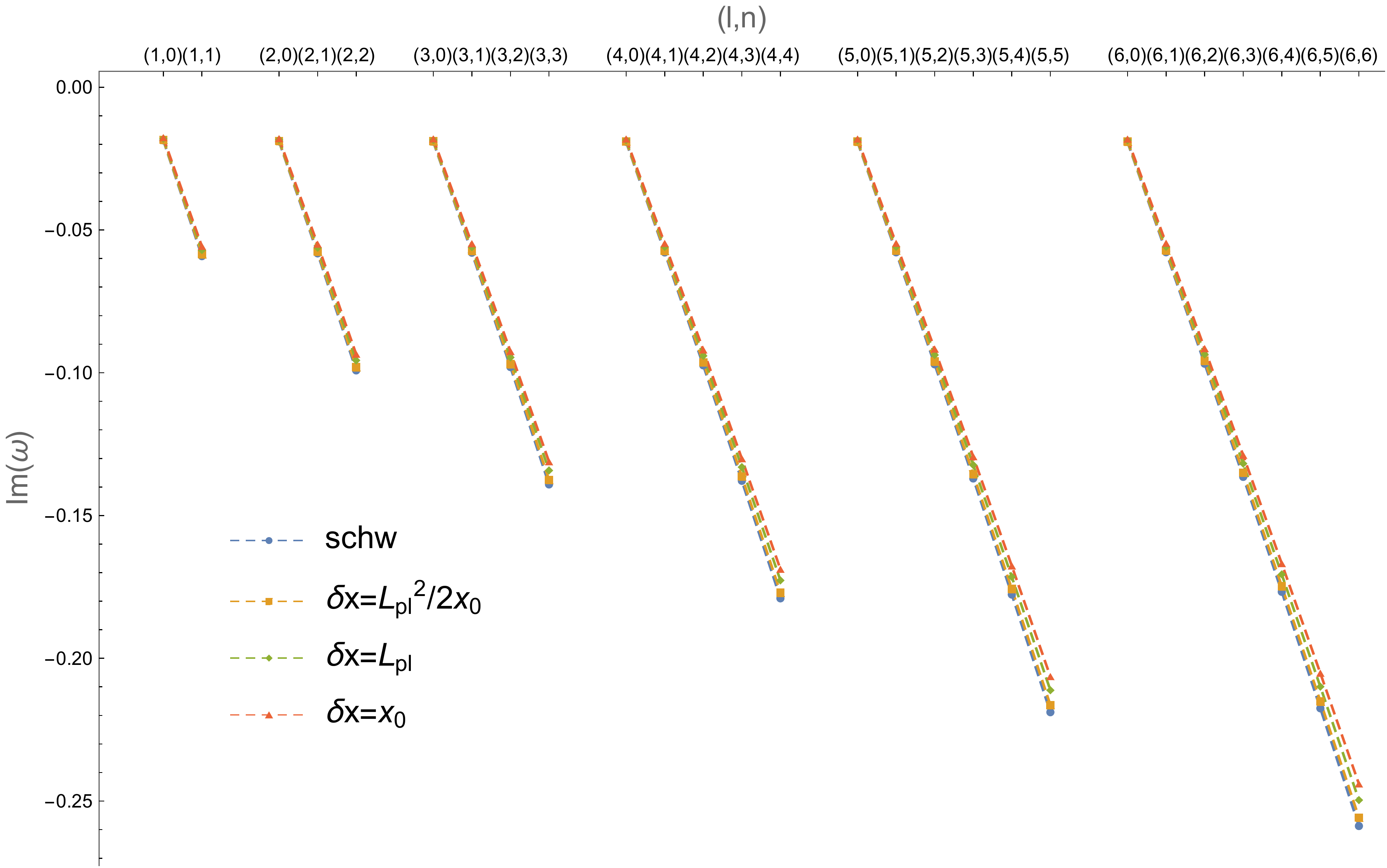}
}
\caption{The real part (top) and the imaginary part (bottom) of the QNMs for the electromagnetic field perturbation are presented. Different markers represent different values of the parameter $\delta x$: the square, diamond, and triangle correspond to $\delta x=\ell_{\rm Pl}^2/2x_0$, $\delta x=\ell_{\rm Pl}$, and $\delta x=x_0$ respectively. The Schwarzschild case is presented by the point. The panels (top and bottom) show how frequencies change with respect to the change of the multipole number $l$ and overtone number $n$. The data used here are from Table \ref{tbe}. }
\label{figew}
\end{figure}

Similarly, we use the WKB approximate method to calculate the QNM frequencies of the quantum black hole. We provide the QNM frequencies for the multipole number $l= 1, 2, 3, 4, 5$ and $6$ for different $\delta x$ in Table \ref{tbe} (which can be seen in the Appendix), and plot QNM frequencies of the quantum black hole for different values of the parameter $\delta x$ as shown in Fig. \ref{figew}.
From there we show that the QNM frequencies of the electromagnetic field perturbations have the same qualitative tendency for the variation of the quantum parameters, the same as the massless scalar field perturbations with $n\neq0$.

\subsection{The axial gravitational perturbations}
Now let us turn to the axial gravitational perturbation of the GOP quantum black hole. The perturbed metric $g_{\mu\nu}$ can be written as \cite{Chandrasekhar:1985kt}
\bq
\label{ax1}
ds^2=-|g_{tt}|dt^2+r^2\sin^2\theta(d\phi-\chi dt-q_{2}dr-q_{3}d\theta)^2+g_{rr}dr^2+r^2d\theta^2,
\eq
where $\chi$, $q_{2}$, and $q_{3}$ are, respectively, the functions of time $t$, radial coordinate $r$, and polar angle $\theta$.
Here we will use a strategy which is similar with what previous work \cite{Bouhmadi-Lopez:2020oia} used, where it assumes that quantum correction be an anisotropic fluid.

In the tetrad formalism, axial perturbations are characterized by the nonvanishing of $\chi$, $q_{2}$, and $q_{3}$. The equations governing these quantities are given by the axial components of $R_{(a)(b)}=0$
\bq
R_{(\phi)(r)}=R_{(\phi)(\theta)}=0,
\eq
we insert for the unperturbed values \eqref{ax1}, and the resulting equations are
\bq
\label{ax2}
\Big[r^2\sqrt{\frac{|g_{tt}|}{g_{rr}}}(q_{2,\theta}-q_{3,r})\Big]_{,r}=r^2\sqrt{\frac{g_{rr}}{|g_{tt}|}}(\chi_{,\theta}-q_{3,t})_{,t}
(\delta R_{\phi \theta}=0),
\eq
\bq
\label{ax3}
\Big[r^2\sqrt{\frac{|g_{tt}|}{g_{rr}}}(q_{3,r}-q_{2,\theta})\sin^3\theta\Big]_{,\theta}=\frac{r^4\sin^3\theta}{\sqrt{|g_{tt}|g_{rr}}}(\chi_{,r}-q_{2,t})_{,t}
(\delta R_{\phi r}=0).
\eq
Then we define
\bq
\mathcal{Q}\equiv r^2\sqrt{\frac{|g_{tt}|}{g_{rr}}}(q_{2,\theta}-q_{3,r})\sin^3\theta e^{-i \omega t}.
\eq
Eliminating $\chi$ from \eqref{ax2} with respect to $r$ and \eqref{ax3} with respect to $\theta$, we obtain
\bq
\label{ax4}
r^4(\sqrt{\frac{|g_{tt}|}{g_{rr}}}\frac{\mathcal{Q}_{,r}}{r^2})_{,r}+\sin^3\theta\sqrt{|g_{tt}|g_{rr}}(\frac{\mathcal{Q}_{,\theta}}{\sin^3\theta})_{,\theta}+\omega^2r^2\sqrt{\frac{g_{rr}}{|g_{tt}|}}\mathcal{Q}=0.
\eq
Here we make the following field decomposition:
\bq
\label{ax5}
\mathcal{Q}=\sum_{\ell m} \varpi_{\ell m}(r)C^{-3/2}_{l+2}(\theta) e^{-i\omega t},
\eq
where $C^{-3/2}_{l+2}(\theta)$ is the Gegenbauer function, satisfied
\bq
\sin^3\theta\frac{d}{d\theta}(\frac1{\sin^3\theta}\frac{d}{d\theta}C(\theta))=-(l+2)(l-1)C(\theta).
\eq
With the substitution \eqref{ax5}, Eq. \eqref{ax4} can be transformed to
\bq
\Big(\sqrt{\frac{|g_{tt}|}{g_{rr}}}\frac{\varpi_{,r}}{r^2}\Big)_{,r}+\omega^2\sqrt{\frac{g_{rr}}{|g_{tt}|}}\frac{\varpi}{r^2}-\frac{\sqrt{|g_{tt}|g_{rr}}}{r^4}(l-1)(l-2)\varpi=0.
\eq
Then we redefine $\psi_{G}\equiv \varpi/r$, bring $x$ back, and use the tortoise radius defined by (\ref{tortx}). The above equation can be written as
\bq
\partial_{r_{\ast}}^2 \psi_{G}+(\omega^2-V_G(x))\psi_{G}=0,
\eq
and the effective potential $V_G(x)$ have the form
\bq
V_G(x)=|g_{tt}|\left[\frac{l(l+1)}{(x+x_0)^2}+\frac{2(g_{xx}^{-1}-1)}{(x+x_0)^2}-\frac1{(x+x_0)\sqrt{|g_{tt}|g_{xx}}}\left(\frac{d}{d x}\sqrt{\frac{|g_{tt}|}{g_{xx}}}\right)\right],
\eq
where the $g_{tt}$, $g_{rr}$ are the same as \eqref{gt} and \eqref{gx}.

\begin{figure}[h!]
\centering
  \includegraphics[width = 0.7\textwidth]{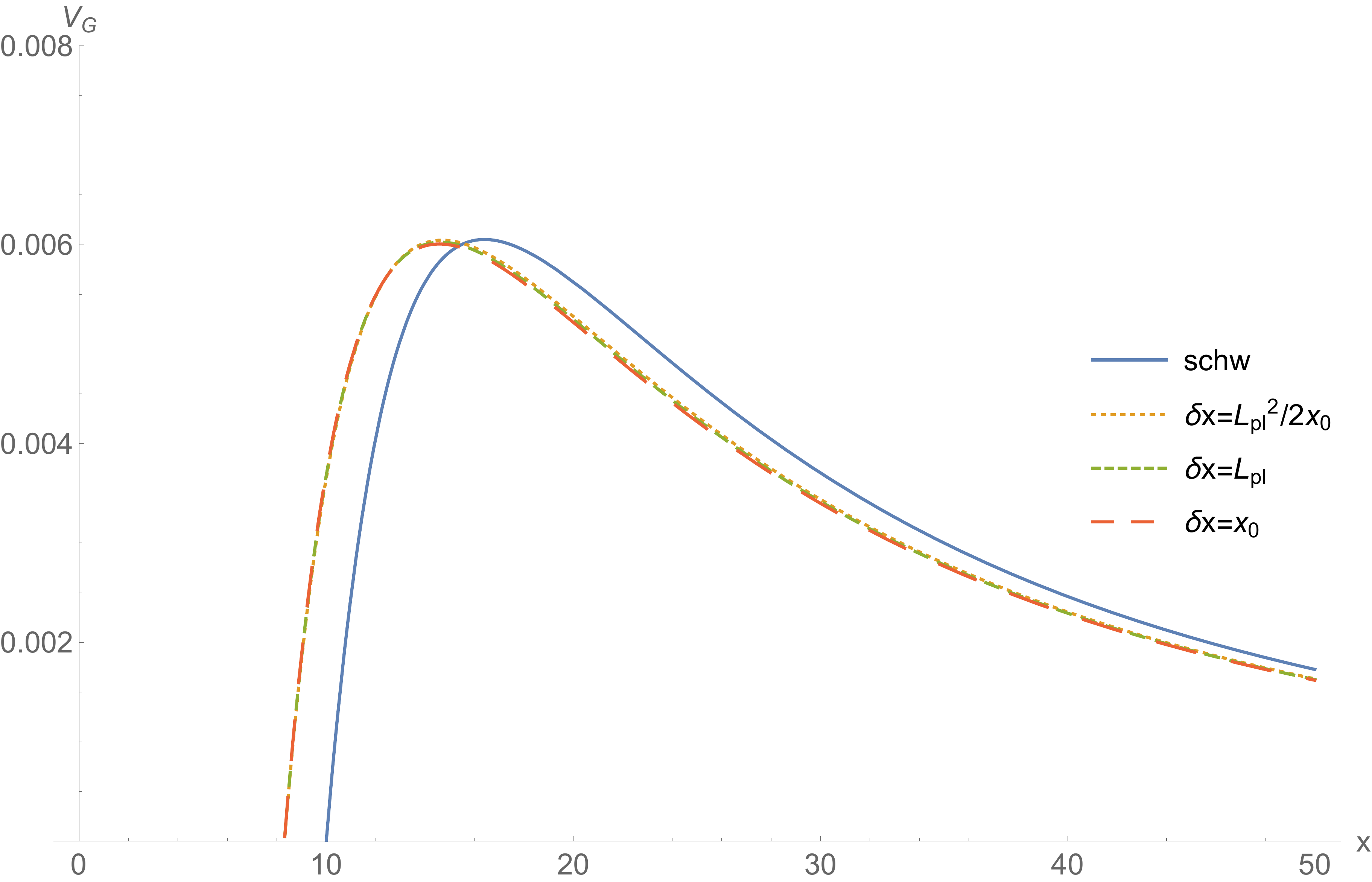}
\caption{The effective potential $V_G(x)$ is shown for different values of parameters. The dotted, short dash, and long dash curves correspond to $\delta x=\ell_{\rm Pl}^2/2x_0$, $\delta x=\ell_{\rm Pl}$, and $\delta x=x_0$ respectively. The potential corresponding to the Schwarzschild solution is presented by the solid curves. Here we applied $M_0=10$ and the multipole number $l=2$.}
\label{figVg}
\end{figure}

\begin{figure}[htb]
{\centering
  \includegraphics[width = 0.7\textwidth]{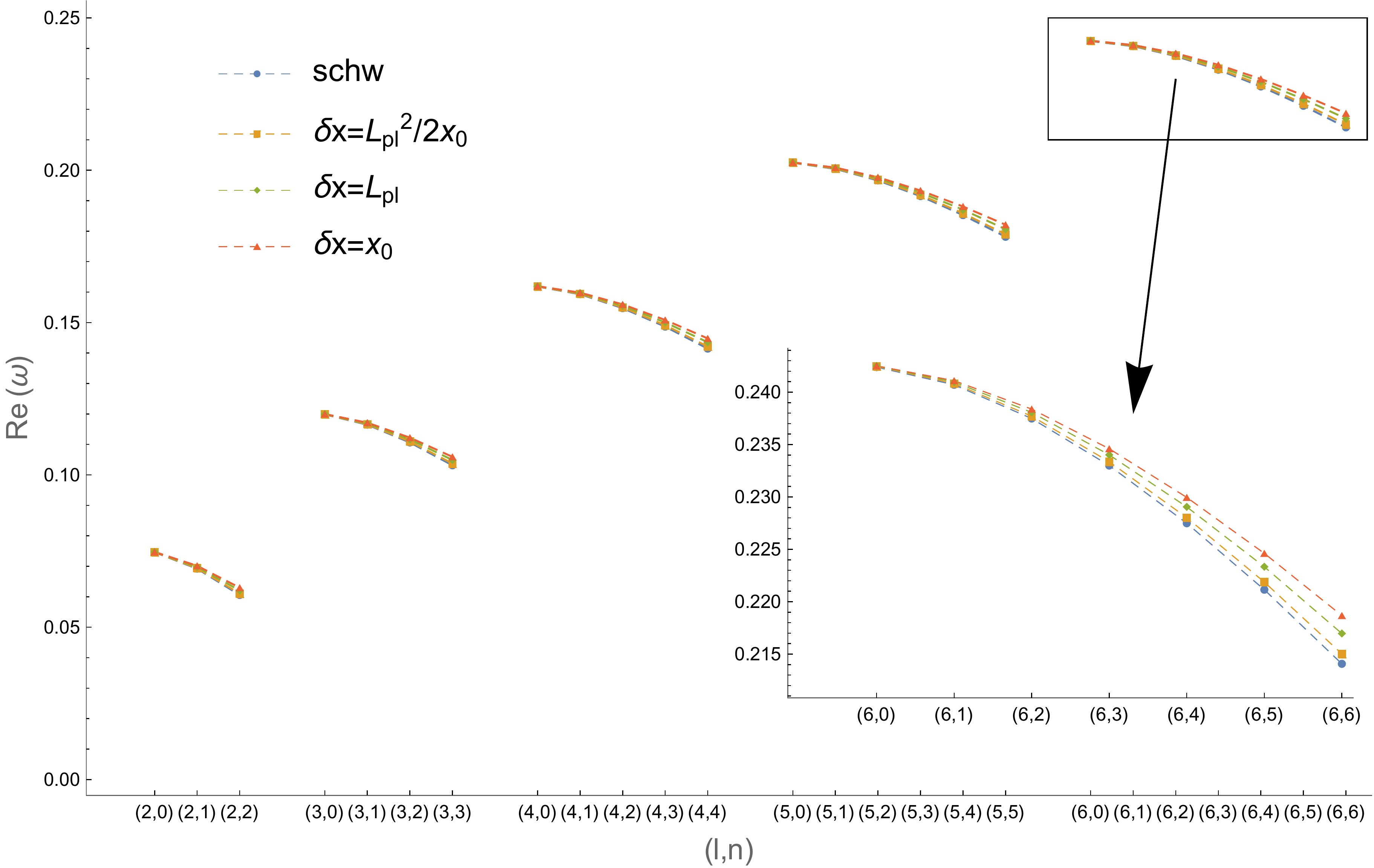}\hspace{0.45cm}
  \includegraphics[width = 0.7\textwidth]{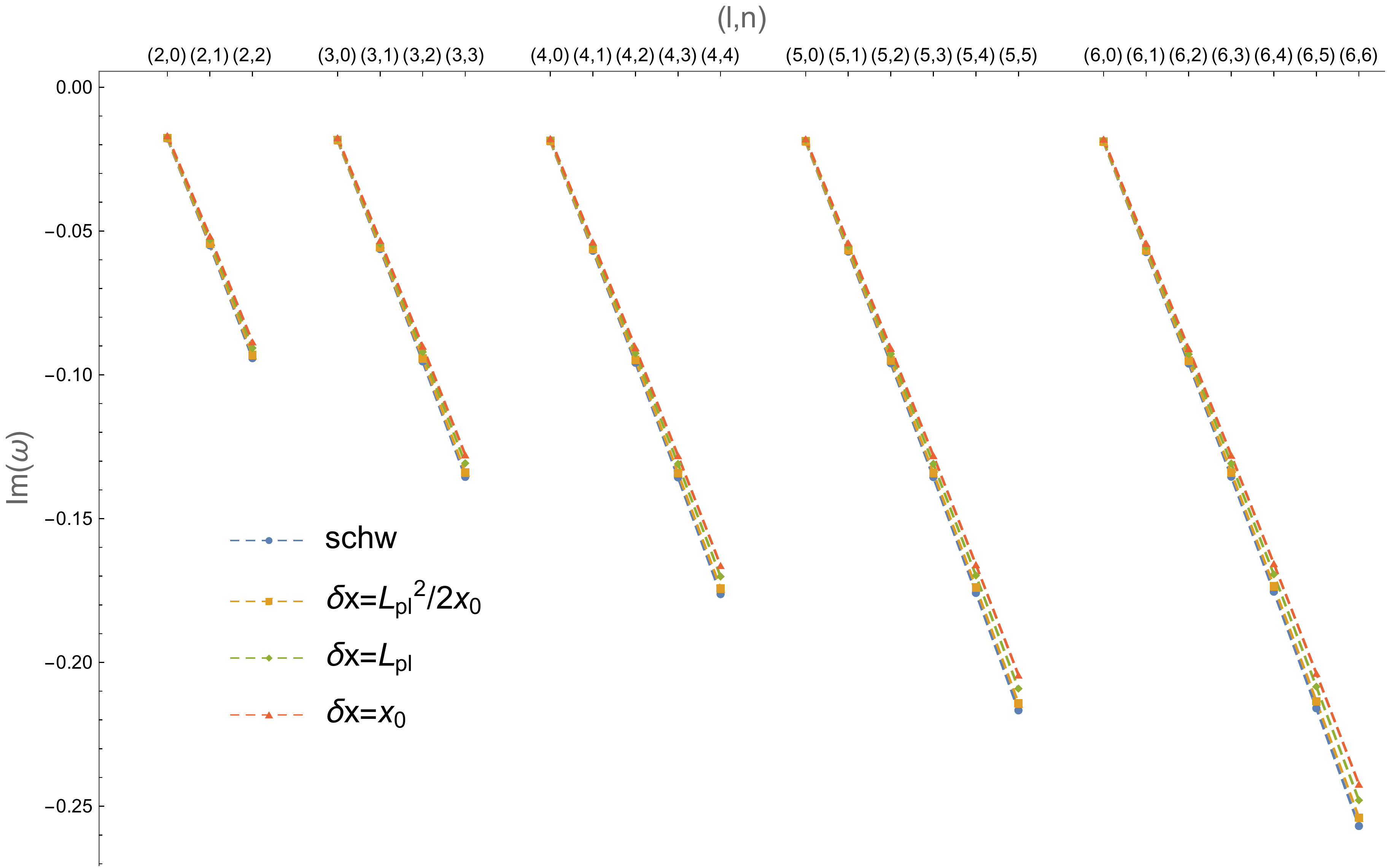}
}
\caption{The real part (top) and the imaginary part (bottom) of the QNMs for the axial gravitational perturbation are presented. Different markers represent different values of the parameter $\delta x$: the square, diamond, and triangle correspond to $\delta x=\ell_{\rm Pl}^2/2x_0$, $\delta x=\ell_{\rm Pl}$, and $\delta x=x_0$ respectively. The Schwarzschild case is presented by the point.  The panels (top and bottom) show how frequencies change with respect to the change of the multipole number $l$ and overtone number $n$. The data used here are from Table \ref{tbg}. }
\label{figgw}
\end{figure}

The effective potential $V_G(x)$ of the quantum black hole with different values of parameters and the Schwarzschild black hole is shown in Fig. \ref{figVg}.
Again, we use the WKB approximate method to calculate the QNM frequencies of the quantum black hole. We provide the QNM frequencies for the multipole number $l=2, 3, 4, 5$ and $6$ for different $\delta x$ in Table \ref{tbg} (which can be seen in the Appendix), and plot QNM frequencies of the quantum black hole for different values of the parameter $\delta x$ in Fig. \ref{figgw}. Once again, we find the QNM frequencies exhibit the qualitative tendency for the variation of quantum parameters, as observed in the scalar (for $n\neq0$) and the electromagnetic (EM) cases.
\subsection{Eikonal limit}
It is well known that the WKB approximate method is more accurate for $l>n$, so it is useful to consider the QNMs in the large multipole number $l$. In other words, we would like to study the Eikonal limit of the QNM expansion. In the limit $\ell \rightarrow \infty$, QNMs can be described as null particles trapped at the unstable circular orbit and slowly leaking out,
in which $\omega_R$, $\omega_I$ depend on the angular velocity at the unstable null geodesic and the instability timescale of the orbit \cite{Cardoso:2008bp}.

In the limit $\ell \rightarrow \infty$, $Q_0$ in \eqref{Q0} satisfies
\bq
2a(x_m)=x_m a'(x_m),
\eq
where $x_m$ is the position at which $V$ has the maximum value. Taking a direct calculation, we find $x_m$ almost coincides with the location of the null circular geodesic $r_c$ in the Eikonal limit. Then, the QNM frequencies calculated by the WKB approximate method is naturally truncated at the second-order,
\bq
\omega_{QNM}=\Omega_c l-i (n+1/2)|\lambda|,
\eq
with
\bq
\label{omega}\Omega_c=\sqrt{\frac{a_c}{{r_c}^2}},
\eq
\bq
\label{lambda}\lambda=\frac1{\sqrt{2}} \sqrt{-\frac{(r_c)^2}{a_c} \left(\frac{\mathrm{d}^2}{\mathrm{dr_{\ast}}^2}\frac{a(x)}{(x+x_0)^2}\right)_{x=r_c}},
\eq
where $\Omega_c$ is the coordinate angular velocity in null geodesics, $\lambda$ is the Lyapunov exponents, and $a_c$ is the function $a$ with $x=r_c$, where $r_c$ is the radius of the null circular geodesic.
Now inserting $a(x)$  into (\ref{omega}) and (\ref{lambda}), we have
\bq
\label{omega1}\Omega_c=\frac{\sqrt{675+64\alpha}}{135 M},
\eq
\bq
\label{lambda1}\lambda=\frac{\sqrt{2278125 (2 \beta +3)-8192\alpha^2(77\beta+120)-77760 \alpha(32\beta +53)}}{2025 M (3+2\beta)\sqrt{5}}.
\eq
From (\ref{omega1}) and (\ref{lambda1}), we see that the real part of $\omega_{QNM}$ in the quantum black hole is larger than that of the Schwarzschild black hole, while $\lambda$ in the quantum black hole is smaller than the one in the Schwarzschild black hole. This is consistent with the previous analysis.

As a byproduct, we can now calculate the black hole shadow. References \cite{Stefanov:2010xz,Jusufi:2019ltj} show that the real part of QNMs is inversely proportional to the shadow radius in the Eikonal limit,
\bq
R_S=\lim \limits_{l\rightarrow\infty}\frac{l}{\omega_R}=\frac{r_c}{\sqrt{a_c}}.
\eq
Therefore, it is straightforward to show that the expression of the shadow radius for the GOP black hole is given by
\bq
R_S=\frac1{\Omega_c}=\frac{135 M}{\sqrt{675+64\alpha}}.
\eq
From this expression, we find that the shadow of the quantum black hole is smaller than that of the Schwarzschild black hole. This is one more effect that the GOP black hole can tell us.

Let us make a brief summary of this section. First, it should be emphasized that the GOP quantum black hole is stable under scalar, EM, and axial gravitational perturbations,  since the imaginary part of the frequency is negative as can be seen in Figs. \ref{figsw}, \ref{figew}, and \ref{figgw}. Second, the QNM frequencies for all these three perturbations exhibit similar behavior. Namely, as we change the quantum parameter away from classical one, the real part of frequency becomes higher and higher, meanwhile, it becomes less and less sensitive to $n$. As a contrast, the imaginary part becomes lower and lower in this process. This phenomenon can be explained in the following way:  quantum geometric effects lead spacetimes to have more complicated structures than those given  in the classical case. As a result, waves (or particles) traveling in such spacetimes   will   interact with the (contaminated) background more frequently, and are more energetic but harder to leak out. This point is also confirmed by the potential function as shown in Figs. \ref{figVs}, \ref{figVe}, and \ref{figVg}. The quantum black hole has higher potential energy compared to the Schwarzschild black hole. Third, from Figs. \ref{figsw}, \ref{figew}, and \ref{figgw} we see that, compared to the Schwarzschild black hole, the real part of the QNM frequencies for the GOP black hole is very insensitive to $n$. It exhibits a platformlike behavior and it is almost separated by an equal interval as increasing $l$. That is to say, with the increase of the multipole number $l$, the increasing range of the real part of the frequencies tends to be a fixed value. The exact explanation is still missing. However, it maybe has relationship with the quantized horizon, which says that the horizon area $A$ of black holes is quantized in units of the Planck area $\ell_{\rm Pl}^2$ as proposed by Bekenstein and Mukhanov\cite{Bekenstein:1974jk, Bekenstein:1995ju}.  According to this picture, the frequency of waves emitted by a black hole is quantized; therefore, there are a series of frequencies with nearly equal intervals.

\section{Conclusion}
In this paper, we consider the extended geometry, physical properties, and the QNMs of the GOP quantum black hole, which adopts the improved dynamics scheme of Chiou $et$ $al$. \cite{Chiou:2012pg}
Within the framework of LQG, the quantum parameters in the GOP model depend on the minimal area gap and the size of the discretization of the physical states. As a result, a spacelike transition surface takes place of the classical singularity. Moreover, quantum effect only dominates near the transition surface, is almost negligible at the event horizon, and the model will recover the classical Schwarzschild space at spatial infinities.

%{em}, {EMT_Shell}

To better understand this quantum black hole, we first show that, due to its reflection symmetry with respect to the  transition surface, as shown in Fig. \ref{throat},  the spacetime  is not smooth across this transition surface, and an infinitely thin shell usually appears [cf. Eqs.(\ref{em}) and (\ref{EMT_Shell})].
 With this extension, we then investigate the physics properties of this quantum black hole, including energy conditions of the effective energy-momentum tensor and curvature scalars. By calculating energy density and pressures at the throat, horizon, and the spatial infinities, respectively, we find that none of the three energy conditions is satisfied, which can be regarded as a consequence of the wormhole structure. In order to explore the quantum effect further, we also study the curvature scalars. Our results show that the values of the curvature scalars are no longer diverging at the throat, even though they are still larger, but all in the order of the Planck scales. In addition, as moving away from the transition surface, we can see that the influence of quantum effects is increasingly negligible, which is what we expect.

In the third part of the paper, we study the perturbations of the GOP quantum black hole and calculate their QNM frequencies. We first considered the perturbation of massless scalar fields, electromagnetic fields, and then the axial gravitational fields, and calculated the QNM frequencies in the Eikonal limit. It shows that the QNM frequencies for all three cases exhibit similar qualitative tendency. That is, as we change the quantum parameter away from its classical value, the real part of the frequency becomes higher and higher, and it is less and less sensitive to $n$, while the damping rate goes lower and lower.
When we discuss the perturbation frequency of the quantum black hole in the Eikonal limit, we find a feature of the quantum black hole, namely, the shadow of the quantum black hole is smaller than that of the classical Schwarzschild black hole.

The deviation between the quantum black hole and the classical one is due to the presence of quantum correction, which is more obvious in smaller black holes. For astrophysical black holes, such as solar-mass black holes, deviation will be very small,  which means that the deviation between the LQG black hole and the classical Schwarzschild black hole become significant only for Planckian black holes.

\section*{ACKNOWLEDGMENTS}

  We would like to thank Jorge Pullin for reading the first draft of our manuscript carefully, and pointing out some inconsistence between our analytic expressions and numerical
  calculations, which led us to identify a bug in our numerical code and correct the errors.  This work is  partially supported by the National Natural Science
  Foundation of China with Grants No. 11975116, and No. 11975203,
  and the Jiangxi Science Foundation for Distinguished Young Scientists under Grant No. 20192BCB23007.

\newpage

\section*{APPENDIX}

Here we present the data of QNM frequencies in three cases in detail in Tables \ref{tbs}, \ref{tbe}, and \ref{tbg}.

\begin{description}
\setlength{\itemsep}{1pt}
\setlength{\itemsep}{1000pt}
\setlength{\itemsep}{1000pt}
  \item[A.] Massless scalar
  \renewcommand{\arraystretch}{0.95}
\begin{table}[htbp]
\caption{The real and imaginary parts of quasinormal frequencies of the scalar field in the GOP black hole background. Here we choose $r_S=10$.}
\begin{center}
\begin{tabular}{|c|c|c|c|c|}
\hline
   $l,n$ &$V_{\text{scalar}}(\text{schwarzschild})$&$V_{\text{scalar}}(\delta x=\ell_{\rm Pl}^2/2x_0) $& $V_{\text{scalar}}(\delta x=\ell_{\rm Pl})$& $V_{\text{scalar}}(\delta x=x_0)$  \\ \hline
0,0 & $0.0209294 - 0.0230394 i $ &$ 0.0208334 - 0.0226666 i$&$0.0206735 - 0.0219302 i$&$0.0205126 - 0.0212631 i$ \\ \hline
1,0 & $0.0582228 - 0.0196003  i $&$0.0582144 - 0.0193703 i$ &$0.058173 - 0.0189021i$&$0.0581291 - 0.0184708 i$  \\ \hline
1,1 & $0.0524424 - 0.0614865i $&$ 0.0525971 - 0.0607069  i$ &$ 0.0529531 - 0.0590997 i$&$0.0532592 - 0.0576278i$ \\ \hline
2,0 & $0.0966422 - 0.019361i $&$0.0966468 - 0.0191471  i$ &$0.0966091 - 0.0187033  i$&$0.0965717 - 0.0182929 i$  \\ \hline
2,1 & $0.0926383 - 0.059162  i $&$ 0.0927664 - 0.0584819 i$ &$0.0930045 - 0.057072  i$&$ 0.0932109 - 0.0557713i$   \\ \hline
2,2 & $0.0863321 - 0.100687 i $&$0.0866458 - 0.099494 i$ &$ 0.0873111 - 0.0970174 i$&$ 0.0878943 - 0.0947357 i$  \\ \hline
3,0 & $0.135041 - 0.0193024 i $&$ 0.135059 - 0.0190925  i$ &$0.135029 - 0.018655  i$&$0.135 - 0.0182499 i$  \\ \hline
3,1 & $0.132083 - 0.0584688 i $&$0.132194 - 0.0578188 i$ &$ 0.132368 - 0.0564658 i$&$ 0.132518 - 0.0552149 i$   \\ \hline
3,2 & $0.126968 - 0.0988236 i $&$0.127235 - 0.0976964 i$ &$0.127753 - 0.0953529 i$&$ 0.128209 - 0.0931888  i$  \\ \hline
3,3 & $0.120436 - 0.140211 i $&$0.120901 - 0.138583 i$ &$ 0.121865 - 0.135198i$&$ 0.122712 - 0.132075 i$  \\ \hline
4,0 & $0.173468 - 0.0192793i $&$ 0.173497 - 0.0190709  i$ &$ 0.173473 - 0.0186358 i$&$0.17345 - 0.0182328i$  \\ \hline
4,1 & $0.171137 - 0.0581798 i $&$0.17124 - 0.0575422  i$ &$ 0.171377 - 0.0562125i$&$0.171495 - 0.054982i$   \\ \hline
4,2 & $0.166898 - 0.0979045 i $&$0.167134 - 0.0968103 i$ &$ 0.167557 - 0.0945318  i$&$0.167928 - 0.0924254 i$  \\ \hline
4,3 & $0.161278 - 0.13853 i $&$0.161685 - 0.136955 i$ &$0.162486 - 0.133677 i$&$ 0.163192 - 0.13065  i$  \\ \hline
4,4 & $0.154614 - 0.179865  i $&$0.155225 - 0.177794 i$ &$ 0.156484 - 0.173485 i$&$0.157591 - 0.169508i$  \\ \hline
5,0 & $0.211914 - 0.0192678 i $&$ 0.211953 - 0.0190602  i$ &$ 0.211934 - 0.0186262i$&$0.211914 - 0.0182243i$  \\ \hline
5,1 & $0.209994 - 0.0580329 i $&$0.210096 - 0.0574017 i$ &$0.210208 - 0.0560836 i$&$0.210305 - 0.0548634 i$   \\ \hline
5,2 & $0.206402 - 0.0973913 i $&$0.206617 - 0.0963161  i$ &$ 0.206972 - 0.094074 i$&$ 0.207284 - 0.092  i$  \\ \hline
5,3 & $0.201495 - 0.137488  i $&$0.201861 - 0.135947 i$ &$0.202547 - 0.132736 i$&$ 0.20315 - 0.129769 i$  \\ \hline
5,4 & $0.195578 - 0.178268 i $&$0.196126 - 0.176243i$ &$ 0.197211 - 0.17203 i$&$0.198166 - 0.168137 i$  \\ \hline
5,5 & $0.188826 - 0.219582 i $&$0.189582 - 0.217063i$ &$0.191133 - 0.211825 i$&$ 0.192497 - 0.206986i$   \\ \hline
6,0 & $0.250372 - 0.0192612i $&$0.250422 - 0.0190541 i$ &$0.250405 - 0.0186208 i$&$ 0.250388 - 0.0182194  i$  \\ \hline
6,1 & $0.248742 - 0.0579483 i $&$0.248844 - 0.0573208 i$ &$0.248939 - 0.0560093i$&$ 0.249021 - 0.0547949  i$  \\ \hline
6,2 & $0.245637 - 0.0970788  i $&$0.245838 - 0.0960153 i$ &$0.246144 - 0.0937955  i$&$0.246412 - 0.0917413 i$  \\ \hline
6,3 & $0.241303 - 0.136807 i $&$0.241639 - 0.135288i$ &$0.242237 - 0.132122i$&$0.242763 - 0.129195 i$  \\ \hline
6,4 & $0.235987 - 0.177148 i $&$0.236487 - 0.175158i$ &$0.237442 - 0.171013 i$&$0.238284 - 0.167181 i$  \\ \hline
6,5 & $0.229872 - 0.218028 i $&$0.23056 - 0.215553 i$ &$0.23193 - 0.210402  i$&$0.233136 - 0.205643 i$   \\ \hline
6,6 & $0.223057 - 0.259333i $&$0.223958 - 0.256366i$ &$0.225799 - 0.250192i$&$ 0.227419 - 0.24449i$   \\ \hline
\end{tabular}
\end{center}
\label{tbs}
\end{table}
 \item[B.]
  Electromagnetic field\renewcommand{\arraystretch}{0.95}
\begin{table}[htb]
\caption{The real and imaginary parts of quasinormal frequencies of the electromagnetic field in the GOP black hole background. Here we choose $r_S=10$.}
\begin{center}
\begin{tabular}{|c|c|c|c|c|}
\hline
   $l,n$ &$V_{\text{electr}}(\text{schwarzschild})$  &$V_{\text{electr}}(\delta x=\ell_{\rm Pl}^2/2x_0)$  & $V_{\text{electr}}(\delta x=\ell_{\rm Pl})$& $V_{\text{electr}}(\delta x=x_0)$  \\ \hline
1,0 & $0.049174 - 0.0186212i $&$ 0.0493082 - 0.0184233  i$ &$0.0495768 - 0.018029  i$&$0.0498188 - 0.0176641 i$  \\ \hline
1,1 & $0.0422617 - 0.059167 i $&$ 0.0425954 - 0.0584693 i$ &$ 0.0433285 - 0.0570471  i$&$ 0.0439833 - 0.0557397i$ \\ \hline
2,0 & $0.0914262 - 0.019013 i $&$ 0.0915113 - 0.0188109i$ &$ 0.0916502 - 0.0183944i$&$ 0.0917758 - 0.0180084i$  \\ \hline
2,1 & $0.0871655 - 0.0581943  i $&$ 0.0873842 - 0.0575472  i$ &$ 0.0878158 - 0.056214  i$&$0.0882 - 0.0549821  i$   \\ \hline
2,2 & $0.0804636 - 0.0991724 i $&$ 0.0808855 - 0.0980331 i$ &$ 0.0817717 - 0.09568 i$&$0.0825567 - 0.093509 i$  \\\hline
3,0 & $0.131347 - 0.0191262  i $&$ 0.131421 - 0.0189222  i$ &$0.131516 - 0.0184984 i$&$0.131602 - 0.0181057   i$  \\ \hline
3,1 & $0.128294 - 0.0579592  i $&$0.128466 - 0.0573263  i$ &$ 0.128771 - 0.0560134 i$&$0.129043 - 0.0547985 i$   \\ \hline
3,2 & $0.123022 - 0.0980115  i $&$0.123356 - 0.0969119 i$ &$0.124017 - 0.0946332  i$&$ 0.124603 - 0.0925275 i$  \\\hline
3,3 & $0.116283 - 0.13911 i $&$0.116825 - 0.13752 i$ &$ 0.117947 - 0.134224 i$&$0.118938 - 0.131181 i$  \\\hline
4,0 & $0.170604 - 0.019173 i $&$0.170676 - 0.0189682 i$ &$ 0.170749 - 0.0185413 i$&$0.170815 - 0.0181458 i$  \\ \hline
4,1 & $0.168228 - 0.0578679 i $&$0.168377 - 0.0572407i$ &$0.168613 - 0.0559353 i$&$0.168823 - 0.0547269 i$   \\ \hline
4,2 & $0.163911 - 0.097401i $&$0.164196 - 0.0963237 i$ &$0.164724 - 0.0940849 i$&$ 0.165192 - 0.0920144 i$  \\ \hline
4,3 & $0.158188 - 0.137845i $&$0.158649 - 0.136293i$ &$0.159563 - 0.13307 i$&$ 0.160371 - 0.130092i$  \\ \hline
4,4 & $0.151394 - 0.179003 i $&$0.152067 - 0.176961i$ &$ 0.153448 - 0.172723 i$&$0.154667 - 0.168808i$  \\ \hline
5,0 & $0.209574 - 0.0191968  i $&$ 0.209649 - 0.0189915 i$ &$ 0.209708 - 0.0185631 i$&$0.209762 - 0.0181662i$  \\ \hline
5,1 & $0.20763 - 0.057823  i $&$0.207768 - 0.0571987 i$ &$ 0.207961 - 0.055897 i$&$0.208132 - 0.0546915 i$   \\ \hline
5,2 & $0.203994 - 0.0970495i $&$0.204247 - 0.0959856 i$ &$ 0.204686 - 0.0937703 i$&$ 0.205075 - 0.0917205i$  \\ \hline
5,3 & $0.199027 - 0.137021 i $&$0.199434 - 0.135495 i$ &$0.200208 - 0.132322 i$&$ 0.200892 - 0.129388 i$  \\ \hline
5,4 & $0.193036 - 0.17768 i $&$0.193629 - 0.175675  i$ &$ 0.194808 - 0.171509 i$&$0.195849 - 0.167659i$  \\ \hline
5,5 & $0.186196 - 0.218875i $&$0.187003 - 0.21638i$ &$ 0.188654 - 0.211198 i$&$ 0.19011 - 0.206411i$   \\ \hline
6,0 & $0.248394 - 0.0192105 i $&$0.248473 - 0.019005i$ &$ 0.248523 - 0.0185756i$&$ 0.248569 - 0.0181779 i$  \\ \hline
6,1 & $0.246749 - 0.0577976 i $&$0.246882 - 0.057175 i$ &$0.247045 - 0.0558752 i$&$0.247189 - 0.0546715 i$  \\ \hline
6,2 & $0.243616 - 0.0968319 i $&$0.243849 - 0.0957765 i$ &$0.244225 - 0.093576i$&$0.244558 - 0.0915393 i$  \\ \hline
6,3 & $0.239245 - 0.136468  i $&$0.239615 - 0.13496i$ &$0.240285 - 0.131822 i$&$ 0.240878 - 0.128918  i$  \\ \hline
6,4 & $0.233882 - 0.176721  i $&$0.234418 - 0.174745 i$ &$0.235449 - 0.170634i$&$0.23636 - 0.166833i$  \\ \hline
6,5 & $0.227712 - 0.217514 i $&$0.228439 - 0.215056 i$ &$ 0.229889 - 0.209947 i$&$0.231169 - 0.205225 i$   \\ \hline
6,6 & $0.220834 - 0.258734i $&$0.221777 - 0.255787i$ &$0.223704 - 0.249661i$&$ 0.225401 - 0.244003 i$   \\ \hline
\end{tabular}
\end{center}
\label{tbe}
\end{table}
    \item[C.]
Axial gravitational \renewcommand{\arraystretch}{0.95}
\begin{table}[htb]
\caption{The real and imaginary parts of quasinormal frequencies of the axial gravitational perturbations field in the GOP black hole background. Here we choose $r_S=10$.}
\begin{center}
\begin{tabular}{|c|c|c|c|c|}
\hline
   $l,n$ &$V_{\text{axial}}(\text{schwarzschild})$  &$V_{\text{axial}}(\delta x=\ell_{\rm Pl}^2/2x_0)$  & $V_{\text{axial}}(\delta x=\ell_{\rm Pl})$& $V_{\text{axial}}(\delta x=x_0)$  \\ \hline
2,0 & $0.0746324 - 0.0178435  i $&$0.074656 - 0.0176393 i$ &$ 0.0746772 - 0.0172264 i$&$0.0747016 - 0.016851  i$  \\ \hline
2,1 & $0.0692035 - 0.0549831i $&$0.069394 - 0.0543252i$ &$ 0.0697934 - 0.05299 i$&$0.0701543 - 0.0517779 i$   \\ \hline
2,2 & $0.0605869 - 0.0942128  i $&$ 0.061046 - 0.0930508 i$ &$0.0620629 - 0.0906759i$&$0.0629708 - 0.0885216i$  \\ \hline
3,0 & $0.119853 - 0.0185457 i $&$0.119897 - 0.0183488  i$ &$0.119934 - 0.0179414   i$&$0.119971 - 0.0175655i$  \\ \hline
3,1 & $0.116471 - 0.0562812  i $&$0.116622 - 0.0556686  i$ &$ 0.116889 - 0.054403i$&$ 0.117127 - 0.0532365 i$   \\ \hline
3,2 & $0.11064 - 0.0953368 i $&$0.110972 - 0.0942699  i$ &$0.111631 - 0.0920673 i$&$0.112212 - 0.0900393i$  \\ \hline
3,3 & $0.103149 - 0.135486 i $&$0.103718 - 0.13394 i$ &$ 0.104894 - 0.130747 i$&$ 0.105925 - 0.127811 i$  \\ \hline
4,0 & $0.16182 - 0.0188342  i $&$0.161872 - 0.0186347  i$ &$ 0.161907 - 0.0182195  i$&$0.16194 - 0.0178355 i$  \\ \hline
4,1 & $0.1593 - 0.0568733i $&$0.159433 - 0.0562616i$ &$0.159637 - 0.0549907i$&$ 0.159821 - 0.0538159 i$   \\ \hline
4,2 & $0.154727 - 0.0957948 i $&$0.155002 - 0.0947427 i$ &$ 0.155512 - 0.0925601i$&$ 0.155965 - 0.0905446 i$  \\ \hline
4,3 & $0.148662 - 0.13566 i $&$0.149124 - 0.134143 i$ &$0.15004 - 0.130998 i$&$0.150849 - 0.128095i$  \\ \hline
4,4 & $0.141443 - 0.176253  i $&$ 0.142131 - 0.174255 i$ &$ 0.143541 - 0.170115 i$&$0.144783 - 0.166297 i$  \\ \hline
5,0 & $0.20245 - 0.0189747  i $&$ 0.20251 - 0.0187732 i$ &$ 0.202541 - 0.018353  i$&$0.20257 - 0.0179641  i$  \\ \hline
5,1 & $0.20043 - 0.0571661 i $&$0.200554 - 0.056553 i$ &$ 0.200722 - 0.0552758i$&$0.200872 - 0.0540941 i$   \\ \hline
5,2 & $0.196653 - 0.0959797 i $&$0.196896 - 0.0949341i$ &$0.197317 - 0.0927591 i$&$ 0.19769 - 0.0907483 i$  \\ \hline
5,3 & $0.191496 - 0.13556 i $&$0.191898 - 0.134059i$ &$ 0.192663 - 0.130941i$&$ 0.19334 - 0.128061 i$  \\ \hline
5,4 & $0.185272 - 0.175839i $&$ 0.185866 - 0.173866i$ &$0.187049 - 0.169771i$&$0.188093 - 0.16599i$  \\ \hline
5,5 & $0.178152 - 0.21666 i $&$0.178969 - 0.214205 i$ &$ 0.180641 - 0.209109  i$&$ 0.182114 - 0.204406 i$   \\ \hline
6,0 & $0.242397 - 0.0190534  i $&$0.242464 - 0.0188507i$ &$ 0.242491 - 0.0184275 i$&$ 0.242516 - 0.0180355  i$  \\ \hline
6,1 & $0.240706 - 0.0573313  i $&$0.240828 - 0.056717 i$ &$0.240969 - 0.0554353 i$&$0.241096 - 0.054249  i$  \\ \hline
6,2 & $0.237488 - 0.0960682 i $&$0.237711 - 0.0950264  i$ &$0.23807 - 0.0928558i$&$0.238388 - 0.0908477 i$  \\ \hline
6,3 & $0.232998 - 0.13542 i $&$0.233362 - 0.133932 i$ &$0.23402 - 0.130834  i$&$ 0.234603 - 0.12797i$  \\ \hline
6,4 & $0.227491 - 0.175399 i $&$0.228024 - 0.173447 i$ &$ 0.229051 - 0.169389i$&$0.229958 - 0.165638i$  \\ \hline
6,5 & $0.221149 - 0.215924 i $&$0.221879 - 0.213496i$ &$ 0.223334 - 0.20845 i$&$0.224618 - 0.20379i$   \\ \hline
6,6 & $0.214073 - 0.256879 i $&$0.215025 - 0.253966i$ &$0.216967 - 0.247916 i$&$ 0.218679 - 0.24233 i$   \\ \hline
\end{tabular}
\end{center}
\label{tbg}
\end{table}
\end{description}


\begin{thebibliography}{199}

%\cite{Singh:2009mz}
\bibitem{Singh:2009mz}
P.~Singh,
Are loop quantum cosmos never singular?
Classical Quantum Gravity \textbf{26}, 125005 (2009).
%doi:10.1088/0264-9381/26/12/125005
%[arXiv:0901.2750 [gr-qc]].
%187 citations counted in INSPIRE as of 07 Jul 2021

%\cite{Ashtekar:2011ni}
\bibitem{Ashtekar:2011ni}
A.~Ashtekar and P.~Singh,
Loop quantum cosmology: A
status report,
 Classical Quantum Gravity \textbf{28}, 213001 (2011).
%doi:10.1088/0264-9381/28/21/213001
%[arXiv:1108.0893 [gr-qc]].
%747 citations counted in INSPIRE as of 07 Jul 2021

%\cite{Ashtekar:2005qt}
\bibitem{Ashtekar:2005qt}
A.~Ashtekar and M.~Bojowald,
Quantum geometry and the Schwarzschild singularity,
Classical Quantum Gravity \textbf{23}, 391 (2006).
%doi:10.1088/0264-9381/23/2/008
%[arXiv:gr-qc/0509075 [gr-qc]].
%237 citations counted in INSPIRE as of 07 Jul 2021

%\cite{Modesto:2005zm}
\bibitem{Modesto:2005zm}
L.~Modesto,
Loop quantum black hole,
Classical Quantum Gravity \textbf{23}, 5587 (2006).
%doi:10.1088/0264-9381/23/18/006
%[arXiv:gr-qc/0509078 [gr-qc]].
%166 citations counted in INSPIRE as of 07 Jul 2021

%\cite{Campiglia:2007pb}
\bibitem{Campiglia:2007pb}
M.~Campiglia, R.~Gambini, and J.~Pullin,
Loop quantization of spherically symmetric midi-superspaces : The Interior problem,
AIP Conf. Proc. \textbf{977}, 52 (2008).
%doi:10.1063/1.2902798
%[arXiv:0712.0817 [gr-qc]].
%62 citations counted in INSPIRE as of 07 Jul 2021

%\cite{Boehmer:2007ket}
\bibitem{Boehmer:2007ket}
C.~G.~Boehmer and K.~Vandersloot,
Loop quantum dynamics of the Schwarzschild interior,
Phys. Rev. D \textbf{76}, 104030 (2007).
%doi:10.1103/PhysRevD.76.104030
%[arXiv:0709.2129 [gr-qc]].
%127 citations counted in INSPIRE as of 07 Jul 2021

%\cite{Chiou:2008eg}
\bibitem{Chiou:2008eg}
D.~W.~Chiou,
Phenomenological dynamics of loop quantum cosmology in Kantowski-Sachs spacetime,
Phys. Rev. D \textbf{78}, 044019 (2008).
%doi:10.1103/PhysRevD.78.044019.
%[arXiv:0803.3659 [gr-qc]].
%23 citations counted in INSPIRE as of 04 Jun 2020

\bibitem{GP08} R. Gambini and J. Pullin, Black Holes in Loop Quantum Gravity: The Complete Space-Time, Phys. Rev. Lett. {\bf  101}, 161301 (2008).

%\cite{Modesto:2008im}
\bibitem{Modesto:2008im}
L.~Modesto,
Semiclassical loop quantum black hole,
Int. J. Theor. Phys. \textbf{49}, 1649 (2010).
%doi:10.1007/s10773-010-0346-x.
%[arXiv:0811.2196 [gr-qc]].
%88 citations counted in INSPIRE as of 04 Jun 2020

%\cite{Chiou:2008nm}
\bibitem{Chiou:2008nm}
D.~W.~Chiou,
Phenomenological loop quantum geometry of the Schwarzschild black hole,
Phys. Rev. D \textbf{78}, 064040 (2008).
%doi:10.1103/PhysRevD.78.064040.
%[arXiv:0807.0665 [gr-qc]].
%35 citations counted in INSPIRE as of 04 Jun 2020

%\cite{Brannlund:2008iw}
\bibitem{Brannlund:2008iw}
J.~Brannlund, S.~Kloster, and A.~DeBenedictis,
The evolution
of lambda black holes in the mini-superspace approximation
of loop quantum gravity,
Phys. Rev. D \textbf{79}, 084023 (2009).
%doi:10.1103/PhysRevD.79.084023.
%[arXiv:0901.0010 [gr-qc]].
%21 citations counted in INSPIRE as of 04 Jun 2020

%\cite{Gambini:2013exa}
\bibitem{Gambini:2013exa}
R.~Gambini and J.~Pullin,
An introduction to spherically symmetric loop quantum gravity black holes,
AIP Conf. Proc. \textbf{1647}, 19 (2015).
%doi:10.1063/1.4913331
%[arXiv:1312.5512 [gr-qc]].
%12 citations counted in INSPIRE as of 07 Jul 2021

%\bibitem{GP13} R. Gambini, J. Pullin,  An introduction to spherically symmetric loop quantum gravity black holes, arXiv:1312.5512.

%\cite{Corichi:2015xia}
\bibitem{Corichi:2015xia}
A.~Corichi and P.~Singh,
Loop quantization of the Schwarzschild interior revisited,
Classical Quantum Gravity \textbf{33}, 055006 (2016).
%doi:10.1088/0264-9381/33/5/055006.
%[arXiv:1506.08015 [gr-qc]].
%57 citations counted in INSPIRE as of 04 Jun 2020

%\cite{Cortez:2017alh}
\bibitem{Cortez:2017alh}
J.~Cortez, W.~Cuervo, H.~A.~Morales-Técotl, and J.~C.~Ruelas,
Effective loop quantum geometry of Schwarzschild interior,
Phys. Rev. D \textbf{95}, 064041 (2017).
%doi:10.1103/PhysRevD.95.064041.
%[arXiv:1704.03362 [gr-qc]].
%11 citations counted in INSPIRE as of 04 Jun 2020

%\cite{Olmedo:2017lvt}
\bibitem{Olmedo:2017lvt}
J.~Olmedo, S.~Saini, and P.~Singh,
From black holes to white
holes: A quantum gravitational, symmetric bounce,
Classical Quantum Gravity \textbf{34}, 225011 (2017).
%doi:10.1088/1361-6382/aa8da8.
%[arXiv:1707.07333 [gr-qc]].
%34 citations counted in INSPIRE as of 04 Jun 2020


\bibitem{CR17} C. Rovelli, Planck stars as observational probes of quantum gravity, Nat. Astron. {\bf1}, 0065 (2017).

\bibitem{AP17} A. Perez, Black holes in loop quantum gravity, Rep. Prog. Phys. {\bf 80}, 126901 (2017).

\bibitem{AOS18a} A. Ashtekar, J. Olmedo, and P. Singh, Quantum Transfiguration of Kruskal Black Holes, Phys. Rev. Lett. {\bf 121}, 241301 (2018).

\bibitem{AOS18b} A. Ashtekar, J. Olmedo, and P. Singh, Quantum extension of the Kruskal spacetime, Phys. Rev. D {\bf 98}, 126003 (2018).

%\cite{Alesci:2019pbs}
\bibitem{Alesci:2019pbs}
E.~Alesci, S.~Bahrami, and D.~Pranzetti,
Quantum gravity predictions for black hole interior geometry,
Phys. Lett. B \textbf{797}, 134908 (2019).
%doi:10.1016/j.physletb.2019.134908.
%[arXiv:1904.12412 [gr-qc]].
%11 citations counted in INSPIRE as of 05 Jun 2020

%\cite{Assanioussi:2019twp}
\bibitem{Assanioussi:2019twp}
M.~Assanioussi, A.~Dapor, and K.~Liegener,
Perspectives on the dynamics in a loop quantum gravity effective description of black hole interiors,
Phys. Rev. D \textbf{101}, 026002 (2020).
%doi:10.1103/PhysRevD.101.026002.
%[arXiv:1908.05756 [gr-qc]].
%9 citations counted in INSPIRE as of 04 Jun 2020

%\cite{Bodendorfer:2019nvy}
\bibitem{Bodendorfer:2019nvy}
N.~Bodendorfer, F.~M.~Mele, and J.~M\"unch,
($b,v$)-type variables for black to white hole transitions in effective loop quantum gravity,
Phys. Lett. B \textbf{819}, 136390 (2021).
%doi:10.1016/j.physletb.2021.136390
%[arXiv:1911.12646 [gr-qc]].
%22 citations counted in INSPIRE as of 07 Jul 2021

\bibitem{MDR19} P. Martin-Dussaud and C. Rovelli, Evaporating black-to-white hole, Classical Quantum Gravity {\bf 36}, 245002 (2019).

\bibitem{BMM18} A. Barrau, K. Martineau, and F. Moulin, A status report on
the phenomenology of black holes in loop quantum gravity:
Evaporation, tunneling to white holes, dark matter and
gravitational waves,
Universe {\bf 4}, 102 (2018).

\bibitem{RMD18} C. Rovelli and P. Martin-Dussaud, Interior metric and ray-tracing map in the firework black-to-white hole transition, Classical Quantum Gravity {\bf 35}, 147002 (2018).

\bibitem{BCDHR18} E. Bianchi, M. Christodoulou, F. D'Ambrosio, H. M Haggard, and C. Rovelli, White holes as remnants: a surprising scenario for the end of a black hole, Classical Quantum Gravity {\bf 35}, 225003 (2018) .

\bibitem{AAN20}  D. Arruga, J. Ben Achour, and K. Noui,
Deformed general
relativity and quantum black holes interior,
Universe \textbf{6}, 39 (2020).
%doi:10.3390/universe6030039.
%[arXiv:1912.02459 [gr-qc]].
%4 citations counted in INSPIRE as of 24 Jun 2020

%\cite{Liu:2020ola}
\bibitem{Liu:2020ola}
C.~Liu, T.~Zhu, Q.~Wu, K.~Jusufi, M.~Jamil, M.~Azreg-Anou, and A.~Wang,
 Shadow and quasinormal modes of a rotating
loop quantum black hole,
Phys. Rev. D \textbf{101}, 084001 (2020).
%doi:10.1103/PhysRevD.101.084001.
%[arXiv:2003.00477 [gr-qc]].
%8 citations counted in INSPIRE as of 24 Jun 2020

%\cite{Joe:2014tca}
\bibitem{Joe:2014tca}
A.~Joe and P.~Singh,
Kantowski-Sachs spacetime in loop
quantum cosmology: Bounds on expansion and shear
scalars and the viability of quantization prescriptions,
Classical Quantum Gravity \textbf{32}, 015009 (2015).
%doi:10.1088/0264-9381/32/1/015009
%[arXiv:1407.2428 [gr-qc]].
%39 citations counted in INSPIRE as of 18 May 2021

%\cite{Agullo:2020hxe}
\bibitem{Agullo:2020hxe}
I.~Agullo, V.~Cardoso, A.~D.~Rio, M.~Maggiore, and J.~Pullin,
Potential Gravitational Wave Signatures of Quantum Gravity,
Phys. Rev. Lett. \textbf{126}, 041302 (2021).
%doi:10.1103/PhysRevLett.126.041302
%[arXiv:2007.13761 [gr-qc]].
%14 citations counted in INSPIRE as of 07 Jul 2021

%\bibitem{Agullo20}  I.Agullo, V. Cardoso, A. del Rio, M. Maggiore, and  J. Pullin, Gravitational-wave signatures of quantum gravity, arXiv:2007.13761.

%\cite{Ashtekar:2020ckv}
\bibitem{Ashtekar:2020ckv}
A.~Ashtekar and J.~Olmedo,
Properties of a recent quantum extension of the Kruskal geometry,
Int. J. Mod. Phys. D \textbf{29}, 2050076 (2020).
%doi:10.1142/S0218271820500765
%[arXiv:2005.02309 [gr-qc]].
%22 citations counted in INSPIRE as of 07 Jul 2021

%\bibitem{AO20} A. Ashtekar, J. Olmedo, Properties of a recent quantum extension of the Kruskal geometry, arXiv:2005.02309.

%\cite{Zhang:2020qxw}
\bibitem{Zhang:2020qxw}
C.~Zhang, Y.~Ma, S.~Song, and X.~Zhang,
Loop quantum Schwarzschild interior and black hole remnant,
Phys. Rev. D \textbf{102}, 041502 (2020).
%doi:10.1103/PhysRevD.102.041502
%[arXiv:2006.08313 [gr-qc]].
%6 citations counted in INSPIRE as of 07 Jul 2021

%\bibitem{ZMSZ20} C. Zhang, Y.-G. Ma, S.-P. Song X.-D. Zhang, Loop quantum Schwarzschild interior and black hole remnant, arXiv:2006.08313.

%\cite{Gambini:2020nsf}
\bibitem{Gambini:2020nsf}
R.~Gambini, J.~Olmedo, and J.~Pullin,
Spherically symmetric loop quantum gravity: Analysis of improved dynamics,
Classical Quantum Gravity \textbf{37}, 205012 (2020).
%doi:10.1088/1361-6382/aba842
%[arXiv:2006.01513 [gr-qc]].
%20 citations counted in INSPIRE as of 07 Jul 2021

%\bibitem{GOP20} R. Gambini, J. Olmedo, J. Pullin, Spherically symmetric loop quantum gravity: analysis of improved dynamics, arXiv:2006.01513.

%\cite{Kelly:2020uwj}
\bibitem{Kelly:2020uwj}
J.~G.~Kelly, R.~Santacruz, and E.~Wilson-Ewing,
Effective loop quantum gravity framework for vacuum spherically symmetric spacetimes,
Phys. Rev. D \textbf{102}, 106024 (2020).
%doi:10.1103/PhysRevD.102.106024
%[arXiv:2006.09302 [gr-qc]].
%17 citations counted in INSPIRE as of 07 Jul 2021

%\cite{Giesel:2021dug}
\bibitem{Giesel:2021dug}
K.~Giesel, B.~F.~Li, and P.~Singh,
Non-singular quantum
gravitational dynamics of an LTB dust shell model: The role
of quantization prescriptions, arXiv:2107.05797.
%0 citations counted in INSPIRE as of 24 Jul 2021

%\cite{Garcia-Quismondo:2021xdc}
\bibitem{Garcia-Quismondo:2021xdc}
A.~Garc\'\i{}a-Quismondo and G.~A.~M.~Marug\'an,
 Exploring
alternatives to the Hamiltonian calculation of the Ashtekar-Olmedo-Singh black hole solution, Front. Astron. Space
Sci. {\bf8}, 701723 (2021).
%0 citations counted in INSPIRE as of 24 Jul 2021

%\bibitem{KSWe20}  J. G. Kelly, R. Santacruz, E. Wilson-Ewing, Effective loop quantum gravity framework for vacuum spherically symmetric space-times, arXiv:2006.09302.

\bibitem{Thiemann08} T. Thiemann, {\em Modern Canonical Quantum General Relativity} (Cambridge University Press, Cambridge, England, 2008).

%\cite{Ashtekar:2002sn}
\bibitem{Ashtekar:2002sn}
A.~Ashtekar, S.~Fairhurst, and J.~L.~Willis,
Quantum gravity, shadow states, and quantum mechanics,
Classical Quantum
Gravity \textbf{20}, 1031 (2003).
%doi:10.1088/0264-9381/20/6/302
%[arXiv:gr-qc/0207106 [gr-qc]].
%235 citations counted in INSPIRE as of 02 Jul 2021

%\cite{Ashtekar:2003hd}
\bibitem{Ashtekar:2003hd}
A.~Ashtekar, M.~Bojowald, and J.~Lewandowski,
Mathematical structure of loop quantum cosmology,
Adv. Theor. Math. Phys. \textbf{7}, 233 (2003).
%doi:10.4310/ATMP.2003.v7.n2.a2
%[arXiv:gr-qc/0304074 [gr-qc]].
%607 citations counted in INSPIRE as of 02 Jul 2021

%\cite{Corichi:2007tf}
\bibitem{Corichi:2007tf}
A.~Corichi, T.~Vukasinac, and J.~A.~Zapata,
Polymer
quantum mechanics and its continuum limit,
Phys. Rev. D \textbf{76}, 044016 (2007).
%doi:10.1103/PhysRevD.76.044016
%[arXiv:0704.0007 [gr-qc]].
%127 citations counted in INSPIRE as of 02 Jul 2021

\bibitem{Ashtekar20} A. Ashtekar, Black hole evaporation: A perspective from
loop quantum gravity, Universe {\bf 6}, 21 (2020).

\bibitem{BMM19} N. Bodendorfer, F.M. Mele, and J. M\"unch,
Effective
quantum extended spacetime of polymer Schwarzschild
black hole, Classical Quantum Gravity {\bf 36}, 195015 (2019).

%\cite{Bodendorfer:2019jay}
\bibitem{BMM20}
N.~Bodendorfer, F.~M.~Mele, and J.~M\"unch,
Mass and
horizon Dirac observables in effective models of quantum
black-to-white hole transition, Classical Quantum Gravity \textbf{38}, 095002 (2021).
%doi:10.1088/1361-6382/abe05d
%[arXiv:1912.00774 [gr-qc]].
%27 citations counted in INSPIRE as of 14 Jul 2021

%\cite{Bouhmadi-Lopez:2020oia}
\bibitem{Bouhmadi-Lopez:2020oia}
M.~Bouhmadi-L\'opez, S.~Brahma, C.~Y.~Chen, P.~Chen, and D.~h.~Yeom,
A consistent model of non-singular Schwarzschild black hole in loop quantum gravity and its quasinormal modes,
J. Cosmol. Astropart. Phys. 07 (2020) 066.
%doi:10.1088/1475-7516/2020/07/066
%[arXiv:2004.13061 [gr-qc]].
%15 citations counted in INSPIRE as of 07 Jul 2021

%\bibitem{BL20}  M. Bouhmadi-Lopez, S. Brahma, C.-Y. Chen, P. Chen, D.-h. Yeom, A consistent model of non-singular Schwarzschild black hole in loop quantum gravity and its quasinormal modes,
% arXiv:2004.13061.

 %\cite{Gan:2020dkb}
\bibitem{Gan:2020dkb}
W.~C.~Gan, N.~O.~Santos, F.~W.~Shu, and A.~Wang,
Properties of the spherically symmetric polymer black holes,
Phys. Rev. D \textbf{102}, 124030 (2020).
%doi:10.1103/PhysRevD.102.124030
%[arXiv:2008.09664 [gr-qc]].
%3 citations counted in INSPIRE as of 03 Jul 2021

%\cite{Gambini:2020qhx}
\bibitem{Gambini:2020qhx}
R.~Gambini, J.~Olmedo, and J.~Pullin,
Loop quantum black
hole extensions within the improved dynamics,
Front. Astron. Space Sci. \textbf{8}, 74 (2021).
%doi:10.3389/fspas.2021.647241
%[arXiv:2012.14212 [gr-qc]].
%0 citations counted in INSPIRE as of 07 Jul 2021

%\cite{Wang:2010iop}
\bibitem{Wang:2010iop}
A.~Wang and N.~O.~Santos,
The hierarchy problem, radion mass, localization of gravity and 4D effective Newtonian potential in string theory on $S^{1}/Z_{2}$,
Int. J. Mod. Phys. A \textbf{25}, 1661 (2010).
%doi:10.1142/S0217751X10047890
%[arXiv:0808.2055 [hep-ph]].
%11 citations counted in INSPIRE as of 27 Jul 2021

%\cite{Hawking:1973uf}
\bibitem{Hawking:1973uf}
S.~W.~Hawking and G.~F.~R.~Ellis,
{\em The Large Scale Structure of Space-Time}
(Cambridge University Press, Cambridge, England, 1973).
%1314 citations counted in INSPIRE as of 19 May 2021


 \bibitem{Mei04}  K.A. Meissner,  Black hole entropy in Loop Quantum Gravity, Classical Quantum Gravity {\bf 21} 5245 (2004). % [arXiv:gr-qc/0407052].

\bibitem{Agullo:2010zz}
I.~Agullo, J.~Fernando Barbero, E.~F.~Borja, J.~Diaz-Polo, and E.~J.~S.~Villasenor,
Detailed black hole state counting in loop quantum gravity,
Phys. Rev. D \textbf{82}, 084029 (2010).
%doi:10.1103/PhysRevD.82.084029
%[arXiv:1101.3660 [gr-qc]].
%60 citations counted in INSPIRE as of 27 Dec 2020
%\cite{Engle:2010kt}

\bibitem{Engle:2010kt}
J.~Engle, K.~Noui, A.~Perez, and D.~Pranzetti,
Black hole
entropy from an SU(2)-invariant formulation of type I
isolated horizons,
Phys. Rev. D \textbf{82}, 044050 (2010).
%doi:10.1103/PhysRevD.82.044050
%[arXiv:1006.0634 [gr-qc]].
%125 citations counted in INSPIRE as of 27 Dec 2020

%\cite{Leaver:1986gd}
\bibitem{Leaver:1986gd}
E.~W.~Leaver,
Spectral decomposition of the perturbation response of the Schwarzschild geometry,
Phys. Rev. D \textbf{34}, 384 (1986).
%doi:10.1103/PhysRevD.34.384
%283 citations counted in INSPIRE as of 16 Apr 2021

%\cite{Jansen:2017oag}
\bibitem{Jansen:2017oag}
A.~Jansen,
Overdamped modes in Schwarzschild-de Sitter and a {\em Mathematica} package for the numerical computation of quasinormal modes,
Eur. Phys. J. Plus \textbf{132}, 546 (2017).
%doi:10.1140/epjp/i2017-11825-9
%[arXiv:1709.09178 [gr-qc]].
%60 citations counted in INSPIRE as of 16 Apr 2021

%\cite{Nollert:1999ji}
\bibitem{Nollert:1999ji}
H.~P.~Nollert,
Topical review: Quasinormal modes: The
characteristic "sound" of black holes and neutron stars,
Classical Quantum Gravity \textbf{16}, R159 (1999).
%doi:10.1088/0264-9381/16/12/201
%617 citations counted in INSPIRE as of 16 Apr 2021

%\cite{Konoplya:2011qq}
\bibitem{Konoplya:2011qq}
R.~A.~Konoplya and A.~Zhidenko,
Quasinormal modes of black holes: From astrophysics to string theory,
Rev. Mod. Phys. \textbf{83}, 793 (2011).
%doi:10.1103/RevModPhys.83.793
%[arXiv:1102.4014 [gr-qc]].
%626 citations counted in INSPIRE as of 16 Apr 2021

%\cite{Berti:2009kk}
\bibitem{Berti:2009kk}
E.~Berti, V.~Cardoso, and A.~O.~Starinets,
Quasinormal modes of black holes and black branes,
Classical Quantum
Gravity \textbf{26}, 163001 (2009).
%doi:10.1088/0264-9381/26/16/163001
%[arXiv:0905.2975 [gr-qc]].
%1079 citations counted in INSPIRE as of 16 Apr 2021

\bibitem{Iyer:1986np}
S.~Iyer and C.~M.~Will,
Black hole normal modes: A WKB
approach. 1. Foundations and application of a higher order
WKB analysis of potential barrier scattering,
Phys. Rev. D \textbf{35}, 3621 (1987).
%doi:10.1103/PhysRevD.35.3621
%479 citations counted in INSPIRE as of 15 Jan 2021

%\cite{Konoplya:2003ii}
\bibitem{Konoplya:2003ii}
R.~A.~Konoplya,
Quasinormal behavior of the d-dimensional Schwarzschild black hole and higher order WKB approach,
Phys. Rev. D \textbf{68}, 024018 (2003).
%doi:10.1103/PhysRevD.68.024018
%[arXiv:gr-qc/0303052 [gr-qc]].
%419 citations counted in INSPIRE as of 13 Jan 2021
%\cite{Iyer:1986np}

%\cite{Daghigh:2020fmw}
\bibitem{Daghigh:2020fmw}
R.~G.~Daghigh, M.~D.~Green, and G.~Kunstatter,
Scalar
perturbations and stability of a loop quantum corrected
Kruskal black hole,
Phys. Rev. D \textbf{103}, 084031 (2021).
%doi:10.1103/PhysRevD.103.084031
%[arXiv:2012.13359 [gr-qc]].
%2 citations counted in INSPIRE as of 07 Jul 2021

%\cite{Chandrasekhar:1985kt}
\bibitem{Chandrasekhar:1985kt}
S.~Chandrasekhar,
{\em The Mathematical Theory of Black Holes}
(Oxford University Press, Oxford 1992).
%344 citations counted in INSPIRE as of 30 Mar 2021.
%Oxford Classic Texts in the Physical Sciences. The Clarendon Press, Oxford University
%Press, New York, 1998. ISBN 0-19-850370-9. Reprint of the 1992 edition.

%\cite{Abramowitz:1980mi}
\bibitem{Abramowitz:1980mi}
M.~Abramowitz and I.~Stegun,
{\em Handbook of Mathematical Functions}
(Dover Publications, Dover 1980).

%\cite{Cardoso:2008bp}
\bibitem{Cardoso:2008bp}
V.~Cardoso, A.~S.~Miranda, E.~Berti, H.~Witek, and V.~T.~Zanchin,
Geodesic stability, Lyapunov exponents and quasinormal modes,
Phys. Rev. D \textbf{79}, 064016 (2009).
%doi:10.1103/PhysRevD.79.064016
%[arXiv:0812.1806 [hep-th]].
%375 citations counted in INSPIRE as of 12 Mar 2021

%\cite{Stefanov:2010xz}
\bibitem{Stefanov:2010xz}
I.~Z.~Stefanov, S.~S.~Yazadjiev, and G.~G.~Gyulchev,
Connection between Black-Hole Quasinormal Modes and Lensing in the Strong Deflection Limit,
Phys. Rev. Lett. \textbf{104}, 251103 (2010).
%doi:10.1103/PhysRevLett.104.251103
%[arXiv:1003.1609 [gr-qc]].
%75 citations counted in INSPIRE as of 17 Mar 2021

%\cite{Jusufi:2019ltj}
\bibitem{Jusufi:2019ltj}
K.~Jusufi,
Quasinormal modes of black holes surrounded by
dark matter and their connection with the shadow radius,
Phys. Rev. D \textbf{101}, 084055 (2020).
%doi:10.1103/PhysRevD.101.084055
%[arXiv:1912.13320 [gr-qc]].
%26 citations counted in INSPIRE as of 17 Mar 2021

%\cite{Bekenstein:1974jk}
\bibitem{Bekenstein:1974jk}
J.~D.~Bekenstein,
The quantum mass spectrum of the Kerr black hole,
Lett. Nuovo Cimento \textbf{11}, 467 (1974).
%doi:10.1007/BF02762768
%444 citations counted in INSPIRE as of 29 Apr 2021

%\cite{Bekenstein:1995ju}
\bibitem{Bekenstein:1995ju}
J.~D.~Bekenstein and V.~F.~Mukhanov,
Spectroscopy of the quantum black hole,
Phys. Lett. B \textbf{360}, 7 (1995).
%doi:10.1016/0370-2693(95)01148-J
%[arXiv:gr-qc/9505012 [gr-qc]].
%386 citations counted in INSPIRE as of 29 Apr 2021

%\cite{Chiou:2012pg}
\bibitem{Chiou:2012pg}
D.~W.~Chiou, W.~T.~Ni, and A.~Tang,
Loop quantization of
spherically symmetric midisuperspaces and loop quantum
geometry of the maximally extended Schwarzschild spacetime, arXiv:1212.1265.
%17 citations counted in INSPIRE as of 14 Jul 2021


\end{thebibliography}
\end{document}